  \documentclass[aps,prb,preprint,groupedaddress,eqsecnum]{revtex4-2}

\usepackage{bm}
\usepackage{amsmath}
\usepackage{amssymb}
\usepackage{graphicx}
\usepackage{xcolor}

\usepackage{bbm}

\usepackage{physics}

\begin{document}
\def\opf{\Omega}
\def\opfcc{\omega}
\def\opfe{{\overline{\opf}}}
\def\khat{\hat{k}}
  \def\qhatfrak{\widehat{\mathbb{Q}}}
\def\phat{\hat{p}}
\def\chat{\hat{c}}
\def\rhat{\hat{r}}
\def\zhat{\hat{0}}
\def\Rhat{\hat{R}}
\def\kvec{{\mathbf k}}
\def\qvec{{\mathbf q}}
\def\qscaleV{\bm{\kappa}}
\def\qscaleS{{\kappa}}
\def\zedvec{{\mathbf z}}
\def\rvec{{\mathbf r}}
\def\Rvec{{\mathbf R}}
\def\udisvec{{\mathbf u}}
\def\udisscaF{U}
\def\udisvecF{{\mathbf U}}
\def\zvec{{\mathbf 0}}
\def\Dim{D}
\def\dim{d}
\def\efe{{\cal H}}
\def\efee{{\efe}_{\text e}}
\def\efem{{\efe}_{\text m}}
\def\pno{N}
\def\vol{V}
\def\ctp{\tau}
\def\hrs{\rm HRS}
\def\lrs{\rm LRS}
\def\cocon{g}
\def\locfrac{Q}
\def\scadist{\Pi}
\def\scat{\theta}
\def\sds{{\cal S}}
\def\sdse{\sds_{\text e}}
\def\sdsm{\sds_{\text m}}
\def\hatpi{\hat{\Pi}}
\def\hatth{\hat{\scat}}
\def\tcomb{\mathbb T}
\def\barexi{\xi_{0}}
  \def\formfun{{\cal P}}
  \def\formfunwt{\widetilde{\formfun}}
  \def\formfunnt{          {\formfun}}
\def\ie{{\it i.e.\/}}
\def\eg{{\it e.g.\/}}
\def\viz{{\it viz.\/}}
\def\via{{\it via\/}}
\def\spacedim{D}
\def\repsPHY{A}
\def\alphaPHY{a}
\def\repsREP{n}
\def\alphaREP{\alpha}
\def\opfPHY{{\opf}_{\mathrm phy}}
\def\AVtherm{{\rm th}}
\def\AVdisor{{\rm dis}}
\def\AVrepli{{\rm rep}}
\def\myQsum{\bm{\lambda}}
\def\myQQsum{\bm{\Lambda}}
\def\MUvec{\bm{\mu}}
\def\rmsFl{{\cal F}}
\def\rmsCo{{\cal C}}
  \def\MUvecDUM{{\bf m}}
  \def\rmsFlDUM{{\mathbb{F}}}
  \def\rmsCoDUM{{\mathbb{C}}}
\def\ProbCharsLarge{{\mathbb{P}}}
\def\distPHY{{\widetilde{\ProbCharsLarge}}}
\def\Ucorten{{\mathbb{G}}}
\def\mynewhalf{\tfrac{1}{2}}
\newcommand{\overbar}[1]{\mkern 1.5mu\overline{\mkern-1.5mu#1\mkern-1.5mu}\mkern 1.5mu}
\makeatletter
\newcommand{\Vast}{\bBigg@{3}}
\makeatother
\newcommand*{\Scale}[2][4]{\scalebox{#1}{$#2$}}
\def\mybar{\overbar}
\def\cartD{d}
\def\cartDbar{\mybar{d}}
\def\replangle{{\langle\!\langle}}
\def\reprangle{{\rangle\!\rangle_{\AVrepli}}}
\def\reprangleMF{{\rangle\!\rangle_{\AVrepli}^{\rm MF}}}
\def\reprangleGB{{\rangle\!\rangle_{\AVrepli}^{\rm GB}}}
\def\mydistll{{\cal D}}
\def\mydistlltrans{{\cal W}}
\def\mylolen{\xi}
\def\GBlangle{\{\!\{}
\def\GBrangle{\}\!\}}
\def\mynunderstand{red}
\def\loc{L}
\def\unloc{U}
\def\MUvecSep{\bm{\eta}}
\def\temp{T}
\def\Ucorsca{\Gamma}
\def\shiftll{{\cal{L}}}
\def\swten{{\mathbb{M}}}
\def\myshiftdistll{\widetilde{\mydistll}}
\newcommand{\myalign}{\vphantom{-}}
\def\myIndMet{{\cal G}}
\def\myScaleFn{{\cal F}}
\def\kBolt{k_{\rm B}}
\def\corrker{{\cal K}}
\title{Universal mesoscale heterogeneity and its\\ spatial correlations
in equilibrium amorphous solids}


\author{Boli Zhou}
\email[]{bzhou@utexas.edu}
\affiliation{Department of Physics, University of Texas at Austin,
Austin, TX 78712, USA}

\author{Ziqi Zhou}
\email[]{ziqi.zhou@stonybrook.edu}
\affiliation{Department of Physics and Astronomy, Stony Brook University,
Stony Brook, NY~11794, USA}

\author{Paul M.~Goldbart}
\email[]{paul.goldbart@stonybrook.edu}
\affiliation{Department of Physics and Astronomy, Stony Brook University,
Stony Brook, NY~11794, USA}


  \date{\today}


\begin{abstract}
Candidates for random network media include (but are not limited to) macroscopic systems consisting of long, flexible macromolecules that have been cross-linked (\ie, permanently chemically bonded) to one another at random so as to form the network. 
Owing to their random architectural structure, the characteristics of the thermal motion of the elements of a random network medium vary randomly from point to point throughout the medium, provided the medium has been cross-linked sufficiently to exhibit the equilibrium amorphous solid state. 
A particular replica-statistical field theory has long been known to capture the essential equilibrium physics of the amorphous solid state and the phase transition that gives rise to it. 
Encoded in the {\it mean\/} value of the field associated with this field theory -- \ie, the order parameter for the transition -- is statistical information about the thermal motions of the constituents: 
(i)~the fraction of the constituents that are spatially localized, and 
(ii)~the heterogeneity of the strength of this localization across the medium. 
Encoded in the field's {\it correlations\/} is more refined information about the thermal motions, \eg: 
(i)~how the localization characteristics of pairs of localized constituents are correlated, 
(ii)~how the correlations between the position fluctuations of the pair are distributed, and 
(iii)~how these descriptors vary with the separation of the pair. 
This statistical information about the amorphous solid state is referred to as its mesoscale heterogeneity.   
Increasingly accurate approximations to the order parameter and field correlations are examined, 
beginning with a mean-field description, 
moving up to the incorporation of gapless (\ie, elastic-displacement) fluctuations, and finally 
\via\ the qualitative examination of how the incorporation of the gapped branches of field fluctuations would further improve accuracy.
Hence, an increasingly accurate set of statistical distributions is obtained, which characterize the mesoscale heterogeneity of equilibrium amorphous solids. 
Along the way, particular attention is paid to the role of the {\it induced measure\/}, 
which results from the nonlinear transformation 
from the fluctuating order-parameter field 
to   the fluctuating elastic displacement fields. 
\end{abstract}%



\maketitle

%
\section{Introduction\label{sec:intro}}
\noindent
The amorphous solid state is a state in which a nonzero fraction of particles have become localized, but their mean (\ie, thermal expectation value of their) positions lack any evident long-ranged periodicity. In contrast with their crystalline counterparts, amorphous solids are fundamentally heterogeneous. 
Their physical characteristics -- such as the spatial extent of the thermal motion of their constituent atoms or the correlations amongst such motions -- vary randomly from region to region in a particular sample and also from sample to sample.
The aim of the present Paper is to examine how statistical information about the heterogeneity of a particular class of amorphous solids, \viz, the {\it equilibrium\/} amorphous solids, is encoded in the natural (fluctuating) order-parameter field that characterizes such solids and to decode that information, thus shedding light on the heterogeneity of equilibrium amorphous solids. The results that we obtain depend solely on the structure of this order-parameter field and the form of the Landau-Wilson effective Hamiltonian that governs this field and, hence, governs the equilibrium properties of the amorphous solid state. To the extent that the effective Hamiltonian is universal -- \ie, is fixed in form by arguments of symmetry and length-scales -- its consequences for the heterogeneity of equilibrium amorphous solids are universal.

Several of the questions addressed in the present Paper were initially explored in Refs.~\cite{MGZ-2004,GMZ-2004}. However, subsequent advances, most notably to do with our understanding of the character of the Goldstone branch of low-energy excitations~\cite{NewGoldstone}, suggest that it is worth revisiting and amplifying the exploration of heterogeneity in equilibrium amorphous solids and, in particular, its spatial correlations, beyond the considerations of Refs.~\cite{MGZ-2004,GMZ-2004}.

We note that the work reported here is strictly pertinent only to physical systems in which there is an unambiguous, large separation between the short time-scales characterizing the relaxation of equilibrating freedoms and the long time-scales over which constraints restrict motion and, specifically, inhibit macroscopic crystallization. Thus, we expect the present work to be directly relevant to vulcanized matter (such as rubber and chemical gels) and certain covalently bonded atomic or molecular glasses; 
see Refs.~\cite{D+Ephtl-1976,B+E-1980} and, 
\eg, the discussions in Refs.~\cite{GCZaip-1996,PMGjop-2000}. 
However, we do not think it unreasonable to expect at least indirect relevance to media such as molecular glasses, for which no such separation of time-scales has been identified and therefore to which the application of equilibrium statistical mechanics is more a matter of optimism than of reason; see, \eg, Ref.~\cite{PMGjop-2000}.

The present Paper is organized as follows. 
In Sec.~\ref{sec:physforms} we postulate how the physical order parameter and correlator involving the positions of the particles are determined by the statistical distribution of various parameters that characterize the heterogeneity of the thermal motion of localized particles. 
In Sec.~\ref{sec:repid}, we present the matching of the physical order parameter and physical correlator with their counterparts arising from a replica field-theory description of the system. We also extract the heterogeneity distribution in terms of field-theoretic quantities by means of this matching. 
In Sec.~\ref{sec:reptheory}, we put together results obtained using the replica field theory and use them to compute the physical heterogeneity distribution and understand its implications. 
In Sec.~\ref{sec:conclude}, we give some concluding remarks. 

\section{Key physical diagnostics:
order parameter and correlator
\label{sec:physforms}}
\noindent
The systems we consider comprise a thermodynamically large number $\pno$ of particles whose instantaneous positions are specified by the $\pno$ $\spacedim$-dimensional position-vectors $\Rvec_{j}\,$, indexed by $j=1,\ldots,\pno$ within a macroscopic $\spacedim$-dimensional volume $\vol$. 
Groups of particles certainly may be strung together
\via\ permanent covalent bonds to form a thermodynamically large number of macromolecules, as is the case for vulcanized polymer systems, but structure such as this is not a requirement.
Our focus is on two key {\it physical\/} diagnostics.
(We use the term {\it physical\/} as a differentiator from the distinct but of course related diagnostics that arise \via\ the replica technique.)\thinspace\
The first diagnostic is what we term the {\it physical order parameter\/}. 
Its Fourier-space form~(\ref{eq:PhyOPfourier}) 
and real-space form~(\ref{eq:PhyOPrs})
are the following two functions of
a positive, integral number $\repsPHY$
of $\spacedim$-dimensional vectors,
either the wave-vectors 
$\{\qvec^{\alphaPHY}\}_{\alphaPHY=1}^{\repsPHY}$ 
[in Eq.(\ref{eq:PhyOPfourier})] 
or the position-vectors $\{\rvec^{\alphaPHY}\}_{\alphaPHY=1}^{\repsPHY}$ 
[in Eq.(\ref{eq:PhyOPrs})]:
\begin{subequations}
\begin{eqnarray}
&&
\bigg[{1\over{\pno}}\sum_{j=1}^{\pno}
\prod_{\alphaPHY=1}^{\repsPHY}
\big\langle
\exp({i\qvec^{\alphaPHY}\cdot\Rvec_{j}})
\big\rangle_\AVtherm\bigg]_\AVdisor,
\label{eq:PhyOPfourier}
\\
&&\bigg[{1\over{\pno}}\sum_{j=1}^{\pno}
\prod_{\alphaPHY=1}^{\repsPHY}
\big\langle\delta(\rvec^{\alphaPHY}-\Rvec_{j})
\big\rangle_\AVtherm\bigg]_\AVdisor.%
\label{eq:PhyOPrs}%
\end{eqnarray}%
\label{eq:twoPHYcors}%
\end{subequations}%
Here, angle brackets $\langle\cdots\rangle_\AVtherm$ indicate averages over thermal motion (\ie, annealed averages) and square brackets $[\cdots]_\AVdisor$ indicate averages over realizations of the cross-linking or other constraint instantiations (\ie, quenched or, equivalently, disorder averages). Note that these Fourier-equivalent versions of the order parameter are each averages (over the $\pno$ particles) of products of thermal averages {\it that pertain to a common particle\/}. We assume the system volume to be hyper-cubic in shape and, for convenience, impose on it periodic boundary conditions; these discretize wave-vectors into a finely spaced hyper-cubic lattice of lattice spacing $2\pi/\vol^{1/\spacedim}$.
As has been discussed in various settings~\cite{overviews}, the physical order parameter is capable of detecting and statistically characterizing the single-particle properties of the amorphous solid state. It works because:
(i)~the factors corresponding to a single particle $j$
[\ie, the factors
$\langle\exp i\qvec\cdot\Rvec_{j}\rangle_\AVtherm$
in Eq.~(\ref{eq:PhyOPfourier})
and the factors
$\langle\delta(\rvec-\Rvec_{j})\rangle_\AVtherm$
in Eq.~(\ref{eq:PhyOPrs})]
each detect whether particle $j$ is localized or not;
(ii)~the presence of more than one factor of
$\langle{\exp i\qvec\cdot\Rvec_{j}}\rangle_\AVtherm$
enables the avoidance of the destructive phase interference -- complete, in the thermodynamic limit -- that renders impotent for such purposes the one-factor case as a detector of the anticipated state, \viz, localization without long-ranged periodicity (cf.~Ref.~\cite{ref:EA-1975} for the original application of this order-parameter idea, to the spin glass state); and
(iii)~the wave-vector dependence provides access to the most basic of the statistical characterizations of heterogeneity, \ie, the localization length heterogeneity. This access follows because the dependence of the physical order parameter on a specific combination of (squared) wave-vectors is Laplace-conjugate to the (disorder-averaged) distribution of root-mean-square (\ie, RMS) localization lengths.

The second key diagnostic is what we term the {\it physical correlator\/}.
Its Fourier-space form~(\ref{eq:twoPHYcorsmom}) and
real-space form~(\ref{eq:twoPHYcorsRS})
are the following two functions of a positive,
integral number $\repsPHY$ of {\it pairs\/} of $\spacedim$-dimensional vectors, 
either
the wave-vectors
$\{\qvec_{1}^{\alphaPHY}\}_{\alphaPHY=1}^{\repsPHY}$ and
$\{\qvec_{2}^{\alphaPHY}\}_{\alphaPHY=1}^{\repsPHY}$ 
[in Eq.(\ref{eq:twoPHYcorsmom})] 
or the position-vectors
$\{\rvec_{1}^{\alphaPHY}\}_{\alphaPHY=1}^{\repsPHY}$ and
$\{\rvec_{2}^{\alphaPHY}\}_{\alphaPHY=1}^{\repsPHY}\,$ 
[in Eq.(\ref{eq:twoPHYcorsRS})]:
\begin{subequations}
\begin{eqnarray}
&&\bigg[
{1\over{\pno}^{2}}
\sum_{j_{1},j_{2}=1}^{\pno}
\prod_{\alphaPHY=1}^{\repsPHY}
\big\langle
\exp\left({-i\qvec_{1}^{\alphaPHY}\cdot\Rvec_{j_{1}}}\right)\,
\exp\left({i\qvec_{2}^{\alphaPHY}\cdot\Rvec_{j_{2}}}\right)
\big\rangle_\AVtherm
\bigg]_\AVdisor,
\label{eq:twoPHYcorsmom}
\\
&&\bigg[{1\over{\pno}^{2}}
\sum_{j_{1},j_{2}=1}^{\pno}
\prod_{\alphaPHY=1}^{\repsPHY}
\big\langle
\delta(\rvec_{1}^{\alphaPHY}-\Rvec_{j_{1}})\,
\delta(\rvec_{2}^{\alphaPHY}-\Rvec_{j_{2}})
\big\rangle_\AVtherm\bigg]_\AVdisor.
\label{eq:twoPHYcorsRS}
\end{eqnarray}%
\label{eq:twoPHYcorspair}%
\end{subequations}%
As discussed in Refs.~\cite{MGZ-2004,GMZ-2004} and in the present section of this Paper, the physical correlator -- if known -- would deliver a statistical characterization of the heterogeneity of amorphous solid state that is more refined than the single-particle characterization furnished by the physical order parameter alone.
The correlator (not the connected one) certainly subsumes the physical order parameter~\cite{note:cor2op}; but it also has embedded within it a certain joint distribution function for the heterogeneity that describes how {\it various random localization characteristics associated with {\it pairs\/} of particles\/} are correlated with one another, conditioned on the spatial separation of the mean positions of the particles in the pairs.
As with the physical order parameter, the extracting of the statistical heterogeneity information comes \via\ the Laplace-conjugacy between the distributed parameters and various scalar and tensor combinations built from the $\repsPHY$ pairs of wave-vector arguments.  A straightforward extension of the present discussion would show that higher-(than second)-order correlators encode heterogeneity distributions concerning the thermal motions of triples and larger groupings of particles.

To see how this all works, we address the physical correlator~(\ref{eq:twoPHYcorsmom}) and first consider the contributions to the double sum over the particles coming from pairs of particles in which both are localized.
Let us suppose that the present description seeks to be effective only from macroscopic length-scales down to mesoscopic length-scales (by which we mean length-scales on the order of the typical localization length). Then we are entitled to assume that the localization has a Gaussian profile~\cite{ref:cumulant} and that the factor in Eq.~(\ref{eq:twoPHYcorsmom}) associated with each such pair $(j_{1},j_{2})$ would become
\begin{subequations}%
\begin{eqnarray}
&&
\big\langle
\exp\left({-i\qvec_{1}^{\alphaPHY}\cdot\Rvec_{j_{1}}}\right)\,
\exp\left({ i\qvec_{2}^{\alphaPHY}\cdot\Rvec_{j_{2}}}\right)
\big\rangle_\AVtherm\,\big\vert_{\,j_{1},\,j_{2}\,\in\,{\rm\loc}}
\nonumber
\\
&&
\qquad\qquad=
\exp\left({-i\qvec_{1}^{\alphaPHY}\cdot\MUvec_{j_{1}}}\right)\,
\exp\left({ i\qvec_{2}^{\alphaPHY}\cdot\MUvec_{j_{2}}}\right)\,
\exp\left({(\rmsCo_{j_{1}j_{2}})_{\cartD\cartDbar}\,
(\qvec_{1}^{\alphaPHY})_{\cartD}\,
(\qvec_{2}^{\alphaPHY})_{\cartDbar}}\right)
\nonumber
\\
&&\qquad\qquad\qquad\quad\times
\exp\left({-\mynewhalf(\rmsFl_{j_{1}})_{\cartD\cartDbar}
(\qvec_{1}^{\alphaPHY})_{\cartD}
(\qvec_{1}^{\alphaPHY})_{\cartDbar}}\right)\,
\exp\left({-\mynewhalf(\rmsFl_{j_{2}})_{\cartD\cartDbar}\,
(\qvec_{2}^{\alphaPHY})_{\cartD}\,
(\qvec_{2}^{\alphaPHY})_{\cartDbar}}\right),
\label{eq:onePcor}
\end{eqnarray}
where $j_1,j_2\in{\rm L}$ means that
particles $j_1$ and $j_2$ are both localized,
summations from 1 to $\spacedim$ are implied over the
repeated Cartesian indices $\cartD$ and $\cartDbar$,
$(\qvec)_{\cartD}$ indicates the $\cartD^{\,\rm th}$ 
Cartesian component of the vector $\qvec$
and the distributed localization parameters, \ie,
\begin{eqnarray}
\MUvec_{j}
&\equiv&
\langle
\Rvec_{j}
\,\rangle_\AVtherm\,,
\label{eq:indmean}
\\
(\rmsFl_{j})_{\cartD\cartDbar}
&\equiv&
\langle
(\Rvec_{j}-\MUvec_{j})_{\cartD}\,
(\Rvec_{j}-\MUvec_{j})_{\cartDbar}
\,\rangle_\AVtherm\,,
\label{eq:indfluc}
\\
(\rmsCo_{j_{1}j_{2}})_{\cartD\cartDbar}
&\equiv&
\langle
(\Rvec_{j_{1}}-\MUvec_{{j}_{1}})_{\cartD}\,
(\Rvec_{j_{2}}-\MUvec_{{j}_{2}})_{\cartDbar}
\,\rangle_\AVtherm\,,
\label{eq:indcor}
\end{eqnarray}
respectively characterize the thermal motion: \via\ 
(a)~the mean positions $\MUvec_{j}$ of the individual particles (\ref{eq:indmean}); 
(b)~the position-fluctuation tensors $(\rmsFl_{j})_{\cartD\cartDbar}$ of the individual particles (\ref{eq:indfluc}); and 
(c)~the fluctuation correlation tensors $(\rmsCo_{j_{1}j_{2}})_{\cartD\cartDbar}$ associated with pairs of particles (\ref{eq:indcor}).
When factors~\eqref{eq:onePcor} are multiplied together to form the physical correlator, 
the net phase factor
$\exp\big({
-i\MUvec_{j_{1}}\cdot\sum\nolimits_{\alphaPHY=1}^{\repsPHY}\qvec_{1}^{\alphaPHY}
+i\MUvec_{j_{2}}\cdot\sum\nolimits_{\alphaPHY=1}^{\repsPHY}\qvec_{2}^{\alphaPHY}
}\big)$
generally gives rise to complete destructive interference 
once a summation over thermodynamically large number of particles is performed, 
as a consequence of the absence of crystallinity.  
Such vanishing is avoided only when the associated wave-vectors obey 
$\sum\nolimits_{\alphaPHY=1}^{\repsPHY}\qvec_{1}^{\alphaPHY}=
 \sum\nolimits_{\alphaPHY=1}^{\repsPHY}\qvec_{2}^{\alphaPHY}$.
Hence, after summing over all such pairs of particles the contribution from these pairs to the physical correlator becomes 
\begin{eqnarray}
&&
{1\over{\pno}^{2}}
\sum_{\substack{j_{1}\in{\loc}\\j_{2}\in{\loc}}}\,
\prod_{\alphaPHY=1}^{\repsPHY}
\big\langle
\exp\left({-i\qvec_{1}^{\alphaPHY}\cdot\Rvec_{j_{1}}}\right)\,
\exp\left({i\qvec_{2}^{\alphaPHY}\cdot\Rvec_{j_{2}}}\right)
\big\rangle_\AVtherm
\nonumber 
\\
&&
\qquad\qquad\qquad=
\Big(\delta_{
\sum\limits_{\alphaPHY=1}^{\repsPHY}\,{\qvec_{1}^{\alphaPHY}}, 
\sum\limits_{\alphaPHY=1}^{\repsPHY}\,{\qvec_{2}^{\alphaPHY}}   
}\,\Big)\
{1\over{\pno^{2}}}
\sum_{\substack{j_{1}\in{\loc}\\j_{2}\in{\loc}}}
\exp\left(
-\mynewhalf(\rmsFl_{j_{1}})_{\cartD\cartDbar}
  \sum\limits_{\alphaPHY=1}^{\repsPHY}
(\qvec_{1}^{\alphaPHY})_{\cartD}\,
(\qvec_{1}^{\alphaPHY})_{\cartDbar}
\right)
\nonumber
\\
&&
\qquad\qquad\qquad\qquad\qquad\qquad\times
\exp\left(
-\mynewhalf(\rmsFl_{j_{2}})_{\cartD\cartDbar}
  \sum\limits_{\alphaPHY=1}^{\repsPHY}
(\qvec_{2}^{\alphaPHY})_{\cartD}\,
(\qvec_{2}^{\alphaPHY})_{\cartDbar}
\right)
\nonumber 
\\
&&
\qquad\qquad\qquad\qquad\qquad\qquad\times
\exp\left(
(\rmsCo_{j_{1}j_{2}})_{\cartD\cartDbar}
  \sum\limits_{\alphaPHY=1}^{\repsPHY}
(\qvec_{1}^{\alphaPHY})_{\cartD}\,
(\qvec_{2}^{\alphaPHY})_{\cartDbar}
\right). 
\end{eqnarray}
\end{subequations}

Next, we consider the contribution to the double sum over the particle pairs that comes from pairs in which one particle is localized and the other is delocalized.
To the accuracy required by our focus on length-scales ranging between macroscopic and mesoscopic, we may neglect the consequences of mutual avoidance of the particles in the pair. 
Thus, we may deem the thermal motion of a delocalized particle to be uncorrelated with the thermal motion of its localized counterpart in the pair, and therefore to be found with equal probability in any fixed-volume region of the container.
On the other hand, we take the localized member of the pair to be fluctuating around its mean position with a Gaussian profile.
For this contribution, then, the corresponding factor becomes:
\begin{subequations}
\label{eq:PHYcoronefacLU}
\begin{eqnarray}
&&
\big\langle
\exp\left({-i\qvec_{1}^{}\cdot\Rvec_{j_{1}}}\right)\,
\exp\left({ i\qvec_{2}^{}\cdot\Rvec_{j_{2}}}\right)
\big\rangle_\AVtherm\,\big\vert_{j_1\in{\loc},\,j_2\in{\unloc}}
\nonumber
\\
&&
\qquad\qquad=
\delta_{\zvec,\qvec_{2}^{}}\,
\exp\left({-i\qvec_{1}^{}\cdot\MUvec_{j_{1}}}\right)\,
\exp\left({-\mynewhalf
(\rmsFl_{j_{1}})_{\cartD\cartDbar}\,
(\qvec_{1}^{})_{\cartD}\,
(\qvec_{1}^{})_{\cartDbar}}\,\right),
\end{eqnarray}%
where $j_1\in{\loc}$ and $j_2\in{\unloc}$ indicate that particle $j_1$ is localized and particle $j_2$ is delocalized.
When such factors are multiplied together to form the physical correlator, 
the net phase factor
$\exp\big({-i\sum\nolimits_{\alphaPHY=1}^{\repsPHY}
\qvec_{1}^{\alphaPHY}\cdot\MUvec_{j_{1}}}\big)$
generally gives rise to complete destructive interference 
once a summation over thermodynamically large number of particles is performed -- a consequence of the absence of crystallinity.  
Such vanishing is avoided only when the associated wave-vectors obey $\sum\nolimits_{\alphaPHY=1}^{\repsPHY}\qvec_{1}^{\alphaPHY}={\bf 0}$.
Hence, after summing over all such pairs of particles the contribution from these pairs to the physical correlator becomes
\begin{eqnarray}
&&
{1\over{\pno}^{2}}
\sum_{\substack{j_{1}\in{\loc}\\j_{2}\in{\unloc}}}
\prod_{\alphaPHY=1}^{\repsPHY}
\big\langle
\exp\left({-i\qvec_{1}^{\alphaPHY}\cdot\Rvec_{j_{1}}}\right)\,
\exp\left({i\qvec_{2}^{\alphaPHY}\cdot\Rvec_{j_{2}}}\right)
\big\rangle_\AVtherm
\\
&&
\nonumber
\quad=(1-\locfrac)
\Big(\prod\nolimits_{\alphaPHY=1}^{\repsPHY}
\delta_{\zvec,\qvec_{2}^{\alphaPHY}}\Big)
\Big(\delta_{\zvec,\sum\limits_{\alphaPHY=1}^{\repsPHY}\,
   {\qvec_{1}^{\alphaPHY}}}\,\Big)\,
{1\over{\pno}}\sum_{j_{1}\in{\loc}}
\exp\left({-\mynewhalf(\rmsFl_{j_{1}})_{\cartD\cartDbar}
  \sum\limits_{\alphaPHY=1}^{\repsPHY}
(\qvec_{1}^{\alphaPHY})_{\cartD}
(\qvec_{1}^{\alphaPHY})_{\cartDbar}}\right),
\end{eqnarray}
\end{subequations}
where $\locfrac$ is the fraction of localized particles and,
accordingly, $1-\locfrac$ is the fraction of delocalized particles.
Identical reasoning gives the term resulting from particle pairs for which, instead, 
$j_1\in{\unloc}$ and $j_2\in{\loc}$. 

The remaining situation, in which both particles in the pair in the thermal average factors in Eq.~(\ref{eq:twoPHYcorsmom}) are delocalized, is readily dealt with \via\ the assumption that the particles' avoidance of one another is a microscopic feature that can be neglected at the meso- and longer length-scales,
\ie, where our attention is focused.
Thus, we may take the thermal average of the pair to be given by
\begin{subequations}
\begin{equation}
\big\langle
\exp\left({-i\qvec_{1}^{\alphaPHY}\cdot\Rvec_{j_{1}}}\right)\,
\exp\left({i\qvec_{2}^{\alphaPHY}\cdot\Rvec_{j_{2}}}\right)
\big\rangle_\AVtherm\,\big\vert_{j_{1},\,j_{2}\in{\unloc}}=
\delta_{\zvec,\qvec_{1}^{\alphaPHY}}\,
\delta_{\zvec,\qvec_{2}^{\alphaPHY}}\,,
\end{equation}
\ie, that their thermal motions are independent of one another and uniformly distributed over the container. Accounting for the fact the a fraction $(1-\locfrac)$ of the particles are delocalized, we then have that the contribution to Eq.~(\ref{eq:twoPHYcorsmom}) due to delocalized pairs is given by
\begin{eqnarray}
&&{1\over{\pno}^{2}}
\sum_{j_{1},\,j_{2}\in{\unloc}}
\prod_{\alphaPHY=1}^{\repsPHY}
\big\langle
\exp\left({-i\qvec_{1}^{\alphaPHY}\cdot\Rvec_{j_{1}}}\right)\,
\exp\left({ i\qvec_{2}^{\alphaPHY}\cdot\Rvec_{j_{2}}}\right)
\big\rangle_\AVtherm
=(1-\locfrac)^2
\prod_{\alphaPHY=1}^{\repsPHY}\left(
\delta_{\zvec,\qvec_{1}^{\alphaPHY}}\,
\delta_{\zvec,\qvec_{2}^{\alphaPHY}}\right).
\end{eqnarray}
\end{subequations}

Equipped with the component results just obtained, we may express 
the complete physical correlator~(\ref{eq:twoPHYcorsmom})
[and perforce the physical order parameter~(\ref{eq:PhyOPfourier})]
in the form:
\begin{subequations}
\begin{eqnarray}
\label{eq:phycorproto}
&&\bigg[
{1\over{\pno}^{2}}
\sum_{j_{1},j_{2}=1}^{\pno}
\prod_{\alphaPHY=1}^{\repsPHY}
\big\langle
\exp\left({-i\qvec_{1}^{\alphaPHY}\cdot\Rvec_{j_{1}}}\right)\,
\exp\left({i\qvec_{2}^{\alphaPHY}\cdot\Rvec_{j_{2}}}\right)
\big\rangle_\AVtherm
\bigg]_\AVdisor
\nonumber
\\
&&\quad=
\locfrac^2
\int d(\MUvecDUM_{1},\MUvecDUM_{2},
\rmsFlDUM_{1},\rmsFlDUM_{2},\rmsCoDUM)\,
\ProbCharsLarge(\MUvecDUM_{1},\MUvecDUM_{2},
\rmsFlDUM_{1},\rmsFlDUM_{2},\rmsCoDUM)\,
\exp\left({-i\MUvecDUM_{1}\cdot
\myQsum_{1}
}\right)\,
\exp\left({i\MUvecDUM_{2}\cdot
\myQsum_{2}
}\right)\,
\nonumber
\\
&&\qquad\qquad\qquad\qquad\times
\exp\left(
{-\mynewhalf
(\rmsFlDUM_{1})_{\cartD\cartDbar}\,
(\myQQsum_{11})_{\cartD\cartDbar}
}\right)\,
\exp\left(
{-\mynewhalf
(\rmsFlDUM_{2})_{\cartD\cartDbar}\,
(\myQQsum_{22})_{\cartD\cartDbar}
}\right)\,
\exp\left({
(\rmsCoDUM)_{\cartD\cartDbar}\,
(\myQQsum_{12})_{\cartD\cartDbar}
}\right)\,
\nonumber\\
\noalign{\smallskip}
&&\qquad\qquad+\,\locfrac(1-\locfrac)
\left(\prod\nolimits_{\alphaPHY=1}^{\repsPHY}
\delta_{\zvec,\qvec_{1}^{\alphaPHY}}\right)
\delta_{\zvec,\myQsum_{2}}\;\int d\rmsFlDUM\,\distPHY(\rmsFlDUM)
\exp\left(
{-\mynewhalf
\rmsFlDUM_{\cartD\cartDbar}\,
(\myQQsum_{22})_{\cartD\cartDbar}}\right)
\nonumber
\\
&&\qquad\qquad+\,\locfrac(1-\locfrac)
\left(\prod\nolimits_{\alphaPHY=1}^{\repsPHY}
\delta_{\zvec,\qvec_{2}^{\alphaPHY}}\right)
\delta_{\zvec,\myQsum_{1}}\;\int d\rmsFlDUM\,\distPHY(\rmsFlDUM)
\exp\left(
{-\mynewhalf
\rmsFlDUM_{\cartD\cartDbar}\,
(\myQQsum_{11})_{\cartD\cartDbar}}\right)
\nonumber
\\
&&\qquad\qquad+\,(1-\locfrac)^2
\prod_{\alphaPHY=1}^{\repsPHY}
\left(
\delta_{\zvec,\qvec_{1}^{\alphaPHY}}\,
\delta_{\zvec,\qvec_{2}^{\alphaPHY}}
\right),
\end{eqnarray}%
where the variables
$\{\myQsum_{1},\myQsum_{2},\myQQsum_{11},\myQQsum_{22},\myQQsum_{12}\}$
are certain vector and tensor combinations of the wave-vectors, 
defined below in Eqs.~(\ref{eq:nonrepcombs}).
Observe that we have expressed the first term on the right-hand side 
in terms of a joint probability distribution $\ProbCharsLarge$ 
for the set of localization parameters
$\{\MUvecDUM_{1},\MUvecDUM_{2},\rmsFlDUM_{1},\rmsFlDUM_{2},\rmsCoDUM\}$ 
and the corresponding integration measure 
$d(\MUvecDUM_{1},\MUvecDUM_{2},\rmsFlDUM_{1},\rmsFlDUM_{2},\rmsCoDUM)$. 

We call this distribution the {\it heterogeneity distribution\/};
it is defined as follows:
\begin{eqnarray}
\label{eq:heteroprob}
&&
\ProbCharsLarge(\MUvecDUM_{1},\MUvecDUM_{2},
                \rmsFlDUM_{1},\rmsFlDUM_{2},\rmsCoDUM)
\\
&&\quad\equiv
\bigg[{1\over{\pno}^{2}\locfrac^{2}}
\sum_{j_{1},\,j_{2}\in\loc}
\delta(\MUvecDUM_{1}-\MUvec_{j_{1}})\,
\delta(\MUvecDUM_{2}-\MUvec_{j_{2}})\,
\delta(\rmsFlDUM_{1}-\rmsFl_{j_{1}})\,
\delta(\rmsFlDUM_{2}-\rmsFl_{j_{2}})\,
\delta(\rmsCoDUM    -\rmsCo_{j_{1}j_{2}})\,
\bigg]_\AVdisor,
\nonumber
\end{eqnarray}%
where the various delta functions and corresponding integration measures are the ones appropriate to the (vector or symmetric tensor) arguments of the various delta functions. 
In Eq.~(\ref{eq:phycorproto}), $\ProbCharsLarge$
appears embedded in an integral transform in which its arguments
$\{\MUvecDUM_{1},\MUvecDUM_{2},\rmsFlDUM_{1},\rmsFlDUM_{2},\rmsCoDUM\}$
are exchanged for the quantities 
$\{\myQsum_{1},\myQsum_{2},\myQQsum_{11},\myQQsum_{22},\myQQsum_{12}\}$ 
defined in Eqs.~\eqref{eq:nonrepcombs}. 
Thus we see that \via\ its wave-vector dependence the physical correlator encodes detailed information about the distribution and correlations amongst the localization parameters. 
Observe, too, that we have expressed the second and third terms on the right-hand side of Eq.~(\ref{eq:phycorproto}) in terms of the {\it heterogeneity distribution for single-particle position fluctuations\/},
\begin{eqnarray}
\distPHY(\rmsFlDUM)\equiv
\bigg[{1\over{\pno\locfrac}}
\sum_{j\in{\loc}}
\delta(\rmsFlDUM-\rmsFl_{j})\,
\bigg]_\AVdisor.
\end{eqnarray}
This is not an independent distribution but is, rather, obtainable as a reduction of $\ProbCharsLarge$ over all of its arguments except $\rmsFlDUM_{1}$ (or, equivalently, $\rmsFlDUM_{2}$). It plays the role of the distribution of localization lengths, which features prominently in investigations of the order parameter for the equilibrium amorphous solid state (see, \eg, Refs.~\cite{CGZepl-1994,overviews}).

A key simplification follows from the recognition that the (unreduced) heterogeneity distribution is translationally invariant, depending not individually on the mean positions
$\MUvecDUM_{1}$ and $\MUvecDUM_{2}$
but only on their separation $\MUvecSep$ ($\equiv\MUvecDUM_{2}-\MUvecDUM_{1}$),
so that we have
$\ProbCharsLarge
(\MUvecDUM_{1},\MUvecDUM_{2},\rmsFlDUM_{1},\rmsFlDUM_{2},\rmsCoDUM)=
\vol^{-1}\,
\ProbCharsLarge(\MUvecSep,\rmsFlDUM_{1},\rmsFlDUM_{2},\rmsCoDUM)$, where $\vol$ is the system volume. 
[We use the same symbol, $\ProbCharsLarge$, bearing in mind that, henceforth, only the translationally invariant version (\ie, the one having four arguments) will play a role.]\thinspace\
As a result, the physical correlator becomes
\begin{eqnarray}
&&\bigg[
{1\over{\pno}^{2}}
\sum_{j_{1},j_{2}=1}^{\pno}
\prod_{\alphaPHY=1}^{\repsPHY}
\big\langle
\exp\left({-i\qvec_{1}^{\alphaPHY}\cdot\Rvec_{j_{1}}}\right)\,
\exp\left({i\qvec_{2}^{\alphaPHY}\cdot\Rvec_{j_{2}}}\right)
\big\rangle_\AVtherm
\bigg]_\AVdisor
\label{eq:phycormng}
\nonumber
\\
&&\quad=
\locfrac^2\,\delta_{\myQsum_{1},\myQsum_{2}}\,
\int d(\MUvecSep,
\rmsFlDUM_{1},\rmsFlDUM_{2},\rmsCoDUM)\;
\ProbCharsLarge(\MUvecSep,\rmsFlDUM_{1},\rmsFlDUM_{2},\rmsCoDUM)\,
\exp\left({i\MUvecSep\cdot
\myQsum_{1}
}\right)\,
\nonumber\\
\noalign{\smallskip}
&&\qquad\qquad\qquad\qquad\times
\exp\left(
{-\mynewhalf(\rmsFlDUM_{1})_{\cartD\cartDbar}\,
(\myQQsum_{11})_{\cartD\cartDbar}
}\right)\,
\exp\left({-\mynewhalf(\rmsFlDUM_{2})_{\cartD\cartDbar}\,
(\myQQsum_{22})_{\cartD\cartDbar}
}\right)\,
\exp\left({
(\rmsCoDUM)_{\cartD\cartDbar}\,
(\myQQsum_{12})_{\cartD\cartDbar}
}\right)\,
\nonumber
\\
&&\qquad\qquad+\,\locfrac(1-\locfrac)
\left(\prod\nolimits_{\alphaPHY=1}^{\repsPHY}
\delta_{\zvec,\qvec_{1}^{\alphaPHY}}\right)
\delta_{\zvec,\myQsum_{2}}\;\int d\rmsFlDUM\,\distPHY(\rmsFlDUM)
\exp\left({-\mynewhalf\rmsFlDUM_{\cartD\cartDbar}\,
(\myQQsum_{22})_{\cartD\cartDbar}}\right)
\nonumber
\\
&&\qquad\qquad+\,\locfrac(1-\locfrac)
\left(\prod\nolimits_{\alphaPHY=1}^{\repsPHY}
\delta_{\zvec,\qvec_{2}^{\alphaPHY}}\right)
\delta_{\zvec,\myQsum_{1}}\;\int d\rmsFlDUM\,\distPHY(\rmsFlDUM)
\exp\left({-\mynewhalf\rmsFlDUM_{\cartD\cartDbar}\,
(\myQQsum_{11})_{\cartD\cartDbar}}\right)
\nonumber
\\
&&\qquad\qquad+\,(1-\locfrac)^2
\prod\nolimits_{\alphaPHY=1}^{\repsPHY}
\left(
\delta_{\zvec,\qvec_{1}^{\alphaPHY}}\,
\delta_{\zvec,\qvec_{2}^{\alphaPHY}}
\right).
\end{eqnarray}%
By setting either one or other of the sets of wave-vectors in the physical correlator to zero, 
or by following the argument that leads to Eqs.~\eqref{eq:PHYcoronefacLU}, 
we also arrive at the form of the physical order parameter~(\ref{eq:PhyOPfourier}), \viz, 
\begin{eqnarray}
&&
\bigg[{1\over{\pno}}\sum_{j=1}^{\pno}
\prod_{\alphaPHY=1}^{\repsPHY}
\big\langle
\exp({i\qvec^{\alphaPHY}\cdot\Rvec_{j}})
\big\rangle_\AVtherm\bigg]_\AVdisor
\label{eq:ITtermsone}
\\
&&\qquad\qquad=(1-\locfrac)
\left(\prod\nolimits_{\alphaPHY=1}^{\repsPHY}
\delta_{\zvec,\qvec^{\alphaPHY}}\right)
+\locfrac\,\delta_{\zvec,\myQsum}\;
\int d\rmsFlDUM\,\distPHY(\rmsFlDUM)
\exp\left(
{-\mynewhalf
\rmsFlDUM_{\cartD\cartDbar}\,
\myQQsum_{\cartD\cartDbar}}\right).
\nonumber
\end{eqnarray}
\label{eq:ITterms}%
\end{subequations}%
In the integral transforms, the quantities conjugate to the localization parameters
(\viz, the quantities $\myQsum_{1}$,
$\myQsum_{2}$, $\myQQsum_{11}$,
$\myQQsum_{22}$, and $\myQQsum_{12}$)
depend on the wave-vectors \via\ the following vector and tensor combinations:
\begin{subequations}
\begin{eqnarray}
\myQsum_{1}&\equiv&
\sum\nolimits_{\alphaPHY=1}^{\repsPHY}\qvec_{1}^{\alphaPHY}
\\
\myQsum_{2}&\equiv&
\sum\nolimits_{\alphaPHY=1}^{\repsPHY}\qvec_{2}^{\alphaPHY}
\\
(\myQQsum_{11})_{\cartD\cartDbar}&\equiv&
\sum\nolimits_{\alphaPHY=1}^{\repsPHY}
(\qvec_{1})_{\cartD}^{\alphaPHY}\,
(\qvec_{1})_{\cartDbar}^{\alphaPHY}
\\
(\myQQsum_{22})_{\cartD\cartDbar}&\equiv&
\sum\nolimits_{\alphaPHY=1}^{\repsPHY}
(\qvec_{2})_{\cartD}^{\alphaPHY}\,
(\qvec_{2})_{\cartDbar}^{\alphaPHY}
\\
(\myQQsum_{12})_{\cartD\cartDbar}&\equiv&
\sum\nolimits_{\alphaPHY=1}^{\repsPHY}
(\qvec_{1})_{\cartD}^{\alphaPHY}\,
(\qvec_{2})_{\cartDbar}^{\alphaPHY}\,
\end{eqnarray}%
\label{eq:nonrepcombs}%
 \end{subequations}
Thus we see that {\it knowledge of the physical correlator would indeed furnish information about the distribution of the localization parameters encoded in the heterogeneity distribution\/} $\ProbCharsLarge$.
Observe the following important conceptual point: 
once the wave-vector dependence of the formula for the physical correlator~\eqref{eq:phycormng} has been expressed in terms of the combinations given in Eqs.~(\ref{eq:nonrepcombs})
(and setting aside the terms associated with delocalized particles), 
one no longer resolves the dependence on the number $\repsPHY$ of thermal expectation factors in Eqs.~(\ref{eq:twoPHYcors}). Thus, one can identify a natural continuation away from an integral number of factors, which enables contact to be made with the replica correlator and order parameter.

To summarize what we have discussed in the present section, one can devise two physical diagnostics to detect the amorphous solid state and give a statistical characterization of its heterogeneity. 
One is the physical order parameter, which marks the existence of such a state and provides the fundamental single-particle statistical information, \viz, the fraction of particles that are localized and the distribution of their localization lengths $\distPHY$. 
The other is the physical correlator, which not only incorporates what is provided by physical order parameter but also furnishes a more refined statistical characterization of the heterogeneity of the amorphous solid state \via\ the heterogeneity distribution $\ProbCharsLarge$. 

\section{Replica identifications
\label{sec:repid}}
\noindent
We saw in Sec.~\ref{sec:physforms} that the heterogeneity distribution $\ProbCharsLarge$ is closely related to the physical order parameter and physical correlator. Our strategy for determining $\ProbCharsLarge$ rests on this connection.
It proceeds by using replica theory to obtain what we term 
the {\it replica order parameter\/} $\replangle\Omega\reprangle$ and 
{\it replica correlator\/} $\replangle\Omega^{\ast}\Omega\reprangle$
(which are expectation values involving the fluctuating replica 
order-parameter field $\opf$, as we discuss in detail, below).
Then, from
$\replangle\Omega\reprangle$ and
$\replangle\Omega^{\ast}\Omega\reprangle$
we obtain the
{\it physical order parameter\/}~(\ref{eq:PhyOPfourier}) and
{\it physical correlator\/}~(\ref{eq:twoPHYcorsmom}).
Finally, from the physical order parameter and physical correlator 
we obtain $\ProbCharsLarge$. We now implement this strategy.

As reviewed in detail in Ref.~\cite{GCZaip-1996}, one knows how expectation values involving the replica order-parameter field are related to expectation values involving the replicated microscopic coordinates that feature in the (quenched-randomness-free) effective replica theory that emerges after disorder-averaging. Thus, we have
\begin{subequations}
\begin{eqnarray}
&&
\frac{1}{\pno}\sum_{j=1}^{\pno}
\replangle
\prod\nolimits_{\alphaREP=0}^{\repsREP}
\exp({i
\kvec^{\alphaREP}\cdot\Rvec_{j}^{\alphaREP}})\,
\reprangle=
\replangle\,\Omega(\kvec^{0},\kvec^{1},\dots,
\kvec^{\repsREP})\,\reprangle\,,
\label{eq:repmeaning1}
\\
&&
\label{eq:repmeaning2}
\frac{1}{{\pno}^{2}}\sum_{j_{1},j_{2}=1}^{\pno}
\replangle\,
\prod\nolimits_{\alphaREP=0}^{\repsREP}
\exp({-i
       \kvec_{1}^{\alphaREP}\cdot\Rvec_{j_{1}}^{\alphaREP}})\,
\exp({ i
       \kvec_{2}^{\alphaREP}\cdot\Rvec_{j_{2}}^{\alphaREP}})
\reprangle
\nonumber\\
&&\qquad\qquad\qquad\qquad\qquad\qquad=
\replangle\,
\Omega(\kvec_{1}^{0},\kvec_{1}^{1},\dots,\kvec_{1}^{\repsREP})^{\ast}\,
\Omega(\kvec_{2}^{0},\kvec_{2}^{1},\dots,\kvec_{2}^{\repsREP})
\,\reprangle\,,
\end{eqnarray}
\label{eq:repmeaning}%
\end{subequations}%
where 
$\replangle\cdots\reprangle$ 
means a (microscopic) replica-statistical-mechanical average when the entity being averaged depends on 
replicated particle (\ie, microscopic) coordinates (\ie, $\Rvec_{j}^{\alphaREP}$), 
or it means a (collective) replica-field-theory average when the enclosed entity being averaged 
depends on the field (\ie, $\Omega$). 
Significantly, this pair of replica-theory equations hold only in the formal limit $\repsREP\to 0$ (see also Ref.~\cite{ref:correl-term-add}). 
This means that they have as their independent variables lists 
$(\kvec^{0},\kvec^{1},\dots,\kvec^{\repsREP})$ 
of $(1+\repsREP)$-fold--replicated $\spacedim$-dimensional wave-vectors, 
one such list for Eq.~\eqref{eq:repmeaning1}
and two such lists for Eq.~\eqref{eq:repmeaning2}. 
For convenience, we bundle the $(1+\repsREP)$ vectors into a \lq\lq supervector\rq\rq\ of dimension $(1+\repsREP)\spacedim$ and denote it by $\khat$, so that $\khat\equiv(\kvec^{0},\kvec^{1},\dots,\kvec^{\repsREP})$.
Note the contrast with the
physical order parameter~(\ref{eq:PhyOPfourier}) and
physical correlator~(\ref{eq:twoPHYcorsmom}),
which respectively have as their independent variables 
one or two lists of $\repsPHY$ (and not $1+\repsREP$) wave-vectors, 
where $\repsPHY$ is a strictly positive integer.

So, how can we put the replica-theory equations~(\ref{eq:repmeaning}) to use? 
(i)~Formally, we evaluate the replica field-theory expectation values on the right-hand sides of Eqs.~(\ref{eq:repmeaning}), using statistical field theory for arbitrary values of the independent supervectors $\khat_{1}$ and $\khat_{2}$. However, we characterize the resulting equations as being {\it generically uncomputable\/}, so as to indicate that their terms cannot be evaluated on a computer for generic values of their independent variables, as these are non-integral in number. Then 
(ii)~we extract physical information from these equations by examining them on what we term {\it computable subspaces\/}, \ie, choices of the independent variables that are suitably restricted so that the right-hand sides become not only computable but also physically meaningful, in the sense that the associated left-hand sides then coincide precisely with the physical order parameter and physical correlator, Eqs.~(\ref{eq:twoPHYcorspair}) and (\ref{eq:twoPHYcors}). Thus, in particular, 
(iii)~we arrive at a formula for the left-hand side of Eq.~(\ref{eq:phycorproto}), so that Eq.~(\ref{eq:phycorproto}) enables us to draw conclusions about the heterogeneity distribution $\ProbCharsLarge$.

To follow the procedure just outlined and connect the physical order parameter and correlator to their replica counterparts, we observe that Eqs.~(\ref{eq:repmeaning1}) and (\ref{eq:repmeaning2}) for the replica order parameter and correlator become computable on the families of supervectors $\qhatfrak_{1}$ and $\qhatfrak_{2}$ given by 
\begin{equation}
\qhatfrak_{1}\equiv(\zvec,\qvec_{1}^{1},\dots,\qvec_{1}^{\repsPHY},\zvec,\dots,\zvec) 
\quad{\rm and}\quad
\qhatfrak_{2}\equiv(\zvec,\qvec_{2}^{1},\dots,\qvec_{2}^{\repsPHY},\zvec,\dots,\zvec), 
\label{frakveceq}
\end{equation}
in which exactly $\repsPHY$ (\ie, a strictly positive, integral number of) entries occupy replica 
\lq\lq channels\rq\rq\ $1,\ldots,\repsPHY$. 
However, the replica order parameter and correlator do not only become computable; 
in the $\repsREP\to 0$ limit they also become identical to their physical 
order-parameter and correlator counterparts 
[\ie, Eqs.~(\ref{eq:PhyOPfourier}) and (\ref{eq:twoPHYcorsmom})], 
and hence they provide us with access to the heterogeneity distribution: 
\begin{subequations}
\label{eq:coordPHY}
\begin{eqnarray}
\label{eq:coordPHY1}
&&
\bigg[{1\over{\pno}}\sum_{j=1}^{\pno}
\prod_{\alphaPHY=1}^{\repsPHY}
\big\langle
\exp\bigl({i\qvec^{\alphaPHY}\cdot\Rvec_{j}}\bigr)
\big\rangle_\AVtherm\bigg]_\AVdisor
\nonumber
\\
&&\qquad\qquad\qquad\qquad=
\frac{1}{\pno}\sum_{j=1}^{\pno}
\replangle
\exp\bigl(i\khat\cdot\Rhat_{j}\bigr) \,
\reprangle\big\vert_{\khat=\qhatfrak}
=
\replangle\,\Omega(\khat)\,\reprangle\big\vert_{\khat=\qhatfrak}\,,
\\
\noalign{\medskip}
&& 
\label{eq:REPPHYcor2}
\bigg[
{1\over{\pno}^{2}}
\sum_{j_{1},j_{2}=1}^{\pno}
\prod_{\alphaPHY=1}^{\repsPHY}
\big\langle
\exp\bigl(-i\qvec_{1}^{\alphaPHY}\cdot\Rvec_{j_{1}}\bigr)\,
\exp\bigl( i\qvec_{2}^{\alphaPHY}\cdot\Rvec_{j_{2}}\bigr)
\big\rangle_\AVtherm
\bigg]_\AVdisor
\nonumber
\\
&&\qquad\qquad\qquad\qquad
=
\frac{1}{{\pno}^{2}}\sum_{j_{1},j_{2}=1}^{\pno}
\replangle\,
\exp\bigl({-i\khat_{1}\cdot\Rhat_{j_{1}}}\bigr)\,
\exp\bigl({ i\khat_{2}\cdot\Rhat_{j_{2}}}\bigr)\,
\reprangle\big\vert_{(\khat_{1},\khat_{2})=(\qhatfrak_{1},\qhatfrak_{2})}
\nonumber\\
\noalign{\medskip}
&&\qquad\qquad\qquad\qquad
=
\replangle\,\Omega(\khat_{1})^{\ast}\,\Omega(\khat_{2})
\,\reprangle\big\vert_{(\khat_{1},\khat_{2})=(\qhatfrak_{1},\qhatfrak_{2})}\,, 
\end{eqnarray}%
\label{eq:pairrepexpecs}%
\end{subequations}%
where, again, Eq.~\eqref{eq:REPPHYcor2} holds only when $\khat_{1}\neq\khat_{2}$. 
Note that this association between physical diagnostics and replica theory only holds {\it provided any wave-vector in the zeroth replica is set to zero\/}. This restriction is necessitated by the fact the the zeroth replica is present in the theory as a device for constructing and implementing a suitable model for the distribution of quenched randomness. It is not, therefore, mandated to be associated with the description of freedoms fluctuating thermally in the presence of specified quenched randomness.

In Sec.~\ref{sec:reptheory}, below, we shall see that the results for the right-hand sides of Eqs.~\eqref{eq:repmeaning}, obtained using the replica technique, 
depend on the replicated wave-vectors $(\khat_{1},\khat_{2})$ \via\ certain combinations of these wave-vectors that have {\it the same formal structure\/} as the combinations defined in Eqs.~(\ref{eq:nonrepcombs}).  
Of course, the combinations in Eqs.~(\ref{eq:nonrepcombs}) are built from the positive integer number $\repsPHY$ of physical wave-vectors, whereas the combinations that appear in the results for the right-hand sides of Eqs.~\eqref{eq:repmeaning} are built from the $1+\repsREP$ pairs of replicated wave-vectors {\it in the $\repsREP\to 0$ limit\/}. 
As a result of this structure, 
upon restricting the replicated wave-vectors to the forms given in Eqs.~\eqref{frakveceq} 
[as we do on the right-hand sides of Eqs.~\eqref{eq:pairrepexpecs}], 
we shall find that the dependence on wave-vectors 
of the right-hand sides of Eqs.~\eqref{eq:pairrepexpecs}
has precisely the structure necessary to 
match the right-hand sides of Eqs.~\eqref{eq:ITterms} 
and hence deliver a computable form for (an integral-transformed version of) $\ProbCharsLarge$. 

How do we complete the implementation of our strategy, \ie, use the computable equations to determine the heterogeneity distribution $\ProbCharsLarge$ and the single-particle position-fluctuation distribution $\distPHY$? 
First, we calculate the replica correlator $\replangle\,\Omega(\khat_{1})^{\ast}\,\Omega(\khat_{2})\reprangle$ 
using replica theory at various levels of approximation 
(the details of which will be given in Sec.~\ref{sec:reptheory}). 
Then, we make the restriction to the computable subspace using Eqs.~\eqref{frakveceq}. 
This yields the anticipated structure, \viz, 
an integral transform involving the combinations of vectors~\eqref{eq:nonrepcombs} 
that, furthermore, has the same exponential form as Eq.~\eqref{eq:phycormng}:
\begin{eqnarray}
&&\replangle\,
\Omega(\khat_{1})^{\ast}\,
\Omega(\khat_{2})
\reprangle\big\vert_{(\khat_{1},\khat_{2})=(\qhatfrak_{1},\qhatfrak_{2})}
\label{eq:REPcorform}
\nonumber
\\
&&\quad=
\locfrac^2\,\delta_{\myQsum_{1},\myQsum_{2}}\,
\int d(\rmsFlDUM_{1},\rmsFlDUM_{2},\rmsCoDUM)\;
\formfunnt(\myQsum_{1},
\rmsFlDUM_{1},\rmsFlDUM_{2},\rmsCoDUM)
\nonumber
\\
&&\qquad\qquad\quad
\times
\exp\left(
{-\mynewhalf
(\rmsFlDUM_{1})_{\cartD\cartDbar}\,
(\myQQsum_{11})_{\cartD\cartDbar}
}\,\right)\,
\exp\left({-\mynewhalf
(\rmsFlDUM_{2})_{\cartD\cartDbar}\,
(\myQQsum_{22})_{\cartD\cartDbar}
}\,\right)\,
\exp\left({
(\rmsCoDUM)_{\cartD\cartDbar}\,
(\myQQsum_{12})_{\cartD\cartDbar}
}\,\right)\,
\nonumber
\\
&&\qquad+
\locfrac(1-\locfrac)
\left(\prod\nolimits_{\alphaPHY=1}^{\repsPHY}
\delta_{\zvec,\qvec_{1}^{\alphaPHY}}\right)\delta_{\zvec,\myQsum_{2}}\;\int d\rmsFlDUM\,\formfunwt(\rmsFlDUM)
\exp\left(
{-\mynewhalf
\rmsFlDUM_{\cartD\cartDbar}\,
(\myQQsum_{22})_{\cartD\cartDbar}}\,\right)
\nonumber
\\
&&\qquad+
\locfrac(1-\locfrac)
\left(\prod\nolimits_{\alphaPHY=1}^{\repsPHY}
\delta_{\zvec,\qvec_{2}^{\alphaPHY}}\right)
\delta_{\zvec,\myQsum_{1}}\;
\int d\rmsFlDUM\,\formfunwt(\rmsFlDUM)
\exp\left(
{-\mynewhalf
\rmsFlDUM_{\cartD\cartDbar}\,
(\myQQsum_{11})_{\cartD\cartDbar}}\,\right)
\nonumber
\\
&&\qquad+(1-\locfrac)^2
\left(
\prod\nolimits_{\alphaPHY=1}^{\repsPHY}
\delta_{\zvec,\qvec_{1}^{\alphaPHY}}\,
\delta_{\zvec,\qvec_{2}^{\alphaPHY}}
\right). 
\end{eqnarray}
Here, the distributions 
  $\formfunnt$ 
and 
  $\formfunwt$ 
are merely formal placeholders; concrete results for them will be obtained in Sec.~\ref{sec:reptheory}. 
The first term on the right-hand side of Eq.~\eqref{eq:REPcorform} (\ie, the term of order $\locfrac^{2}$) corresponds to the first term on the right-hand side of Eq.~\eqref{eq:phycormng}, and similarly for the second, third and fourth terms [\ie, the two terms of order $\locfrac(1-\locfrac)$ and the terms of order $\locfrac^{2}$, respectively].
Matching corresponding terms across Eq.~\eqref{eq:REPcorform} and Eq.~\eqref{eq:phycormng} \via\ Fourier transformation, we arrive at the formulas that determine -- in terms of the output of the replica field theory -- the heterogeneity distribution $\ProbCharsLarge$ and, consistent with it, the reduction to single-particle properties $\distPHY({\rmsFlDUM})$: 
\begin{subequations}
\label{eq:distmatching}
\begin{eqnarray}
\ProbCharsLarge(\MUvecSep,
\rmsFlDUM_{1},\rmsFlDUM_{2},\rmsCoDUM)
&=&
{1\over{(2\pi)^{\Dim}}} \int d\myQsum\,
\formfunnt(\myQsum,
\rmsFlDUM_{1},\rmsFlDUM_{2},\rmsCoDUM)\, 
\exp\left({-i\MUvecSep\cdot\myQsum}\right)\,,
\\
\distPHY({\rmsFlDUM})
&=& 
\formfunwt(\rmsFlDUM)\,.
\end{eqnarray}
\end{subequations}
At this stage, we have connected the physical diagnostics that encode statistical information about mesoscale structure in the amorphous solid state to their counterparts in replica field theory. In order to determine these physical diagnostics, we turn to the replica field theory. 
\section{Replica field theory
\label{sec:reptheory}}
Now we compute the replica order parameter~(\ref{eq:repmeaning1}) and replica correlator~(\ref{eq:repmeaning2}) using the framework of replica field theory. 
The material in Secs.~\ref{sec:physforms} and \ref{sec:repid} is rooted in equivalences that are exact, assuming the validity of the replica technique. In the remaining sections we treat the replica field theory at various levels of approximation. (The replica order parameter has been known at the mean-field level of approximation for some time~\cite{CGZepl-1994,overviews}.)

Following the field-theory framework spelled out in Ref.~\cite{GCZaip-1996}, we consider the real-valued replica order-parameter field $\opf$, which depends on the replicated position-vectors 
$\{\rvec^{\alphaREP}\}_{\alphaREP=0}^{\repsREP}$ 
(or, equivalently, the Fourier transform of $\opf$, which depends on the conjugate replicated wave-vectors 
$\{\qvec^{\alphaREP}\}_{\alphaREP=0}^{\repsREP}\,$).
The expectation value and fluctuations of $\opf$ are governed by the effective replica Hamiltonian $\efe_{{\rm e}}$ (see, \eg, Ref.~\cite{PCGZepl-1998}), which is given by: 
\begin{equation}
\efee=
\frac{\pno}{2}\sum_{\khat\in\hrs}
\left(-{a}\ctp+{\barexi^{2}}\,\khat^{2}\right)
\vert\opf(\khat)\vert^{2}
-\pno\cocon
\!\!\!
\sum_{\khat_{1},\khat_{2},\khat_{3}\in\hrs}
\!\!\!
\delta_{\zhat,\khat_{1}+\khat_{2}+\khat_{3}}\,
\opf(\khat_{1})\,\opf(\khat_{2})\,\opf(\khat_{3})\,.
\label{eq:Landau-Wilson}
\end{equation}%
In Eq.~(\ref{eq:Landau-Wilson}), 
the amorphous solidification transition control parameter $\ctp$ is a dimensionless proxy for the density of quenched random constraints (which in the case of vulcanized macromolecules are the chemical cross-links); more precisely, it is a proxy for the amount by which the constraint density exceeds the critical value for the onset of amorphous solidification. 
In addition, $a$ specifies the critical value of the parameter that controls the probability of cross-link formation (and hence specifies the critical density of cross-links) .
Furthermore, $\barexi$ gives the bare size of the entities that the constraints  randomly connect (which in the case of vulcanized macromolecules is proportional to the radius of gyration of the uncross-linked macromolecules); and
$\cocon$ measures the strength of the order-parameter nonlinearity.
We measure energies in units of the thermal energy scale $k_{\rm B}\,T$, where $\temp$ is the temperature and $k_{\rm B}$ is Boltzmann's constant. Furthermore, convenient re-scalings allow us to set $a/2$, $\barexi$ and $g$ to unity.
The indication \hrs\ on a summation over $\khat$ stands for {\it higher replica sector\/}, 
which means that to be included in a summation the corresponding set
$(\kvec^{0},\kvec^{1},\dots,\kvec^{\repsREP})$
must contain at least two replicated wave-vectors that are nonzero. The set of replicated wave-vectors complementary to the \hrs\ set is termed the {\it lower replica sector\/} and indicated \lrs. The freedoms of the field theory associated with the \lrs\ correspond to macroscopic fluctuations of the particle density at some wave-vector, away from spatial homogeneity. As a result of the excluded-volume repulsion between particles, this sector of the field theory is a {\it gapped\/} sector: the free-energy cost does not disperse to zero as the length-scale of the fluctuation diverges. Thus, from the point of view of the critical phenomenology associated with the transition to the amorphous solid state, the \lrs\ freedoms can be omitted from the description: they do not participate in the relevant ordering; nor do they exhibit long-ranged correlations either near the transition or in the ordered state. This is why it is only the freedoms associated with the \hrs\ that are retained in the effective Hamiltonian~(\ref{eq:Landau-Wilson}). 

Our focus here is on the amorphous solid state, and we therefore consider states of the replica theory in which the replica order parameter $\replangle\Omega\reprangle$ is nonzero. 
The notation $\replangle\cdots\reprangle$ is in accordance with the explanation given immediately after Eqs.~\eqref{eq:repmeaning}. In the context of Eq.~\eqref{eq:FTpi}, it means the replica-field-theory average in which expectation values are taken with respect to the un-normalized weight $\exp\left({-\efee}\right)$, \ie, 
\begin{equation}
\replangle\cdots\reprangle
\equiv
{\int{\mathcal D}\Omega\,{\rm e}^{-\efee}\cdots
\over{\int{\mathcal D}\Omega\,{\rm e}^{-\efee}\phantom{\cdots}}}
\label{eq:FTpi}
\end{equation}
where the measure ${\mathcal D}\Omega$ indicates functional integration over all HRS fields, subject to the reality condition $\opf^{\ast}(\khat)=\opf(-\khat)$. 
As we shall need to examine its consequences when we incorporate Goldstone excitations, we give the explicit form of the measure: 
$\mathcal D\Omega\equiv\prod_{\khat\in\hrs}\,d\Re\opf(\khat)\,d\Im\opf(\khat),$ 
where, reflecting the reality condition, the product is taken over only half of the replicated wave-vector space. (For future reference, we note that no explicit metric features in this measure; it is flat.)

Our main focus is on the replica correlator $\replangle\Omega^{\ast}\,\Omega\reprangle$. 
(We could of course equally well focus on the connected version,
$\replangle\Omega^{\ast}\,
           \Omega\reprangle-
\replangle\Omega^{\ast}\reprangle
\replangle\Omega       \reprangle\,$,
which characterizes the correlations between fluctuations of the replica order parameter around its expectation value.)\thinspace\
We adopt a three-level strategy for calculating
$\replangle\Omega^{\ast}\,\Omega\reprangle\,$.
First, we apply mean-field theory. 
Then, we incorporate the effects of the modes contained in the gapless sector of excitations (\ie, the Goldstone branch). 
Finally, we discuss what the implications would be if we were to incorporate the effects of the modes contained in the gapped sector of excitations (\ie, the non-Goldstone branches).

The mean-field solution for the amorphous solidification transition has long been known
(see Refs.~\cite{CGZepl-1994,PCGZepl-1998}),
so here we simply recapitulate its results,
including to fix notation.
The equilibrium expectation value of the replica order parameter $\replangle\Omega\reprangle$ is given, at the mean-field level, by the value of $\Omega$ that makes $\efee$ stationary. For $\ctp\le 0$ it is $0$; for $\ctp>0$ it is given by
\begin{subequations}
\begin{eqnarray}
&&\replangle\,
\Omega(\khat)\,
\reprangle
\stackrel{{\rm MF}}{=}
(1-\locfrac)\,\delta_{\zhat,\khat}+\locfrac\,
\mydistlltrans(\khat^{2})\,
\delta_{\zvec,
\sum\nolimits_{\alphaREP=0}^{\repsREP}{\kvec^{\alphaREP}}}\,,
\label{eq:opformAA}\\
&&\mydistlltrans(L)
\equiv
\int d\mylolen^2\,\mydistll(\mylolen^2)\,
\exp\left({-\mynewhalf\mylolen^{2}L}\right).
\label{eq:opformBB}%
\end{eqnarray}%
\label{eq:opformTot}%
\end{subequations}%
Here, $\locfrac$ and $\mydistll$ are single-particle characteristics of the amorphous solid state, \viz,
the fraction of localized particles and
the normalized distribution of the (squared) localization lengths $\mylolen^2$ of the localized particles, respectively.
(For an analysis of the stability of the amorphous solid state,
see Ref.~\cite{CGZepl-1999}.)\thinspace\
Similarly, within mean-field theory the equilibrium expectation value of the replica correlator $\replangle\Omega^{\ast}\,\Omega\reprangle$
is given by
\begin{eqnarray}
\replangle\,
\Omega(\khat_{1})^{\ast}\,\Omega(\khat_{2})
\,\reprangle
&&
\stackrel{{\rm MF}}{=}
\locfrac^{2}\,
\mydistlltrans\big(\khat_{1}^2\big)\,
\mydistlltrans\big(\khat_{2}^2\big)\,
\delta_{\zvec,\sum\nolimits_{\alphaREP=0}^{\repsREP}{\kvec_{1}^{\alphaREP}}}\,
\delta_{\zvec,\sum\nolimits_{\alphaREP=0}^{\repsREP}{\kvec_{2}^{\alphaREP}}}\,
\nonumber
\\
\nonumber
&&
\qquad
+\,\locfrac(1-\locfrac)\,
\delta_{\zhat,\khat_{1}}
\mydistlltrans\big(\khat_{2}^2\big)\,
\delta_{\zvec,\sum\nolimits_{\alphaREP=0}^{\repsREP}{\kvec_{2}^{\alphaREP}}}\,
\\
\nonumber
&&
\qquad
+\,\locfrac(1-\locfrac)\,
\delta_{\zhat,\khat_{2}}
\mydistlltrans\big(\khat_{1}^2\big)\,
\delta_{\zvec,\sum\nolimits_{\alphaREP=0}^{\repsREP}{\kvec_{1}^{\alphaREP}}}\,
\\
&&
\qquad
+\,(1-\locfrac)^2\,
\delta_{\zhat,\khat_{1}}\delta_{\zhat,\khat_{2}}\,.
\label{eq:corformMF}%
\end{eqnarray}%
By following the steps presented in
Eqs.~\eqref{eq:REPcorform} and \eqref{eq:distmatching},
we arrive at the following mean-field results for the (un-reduced and reduced) heterogeneity distributions:
\begin{subequations}
\label{eq:distMF}
\begin{eqnarray}
\ProbCharsLarge(\MUvecSep,
\rmsFlDUM_{1},\rmsFlDUM_{2},\rmsCoDUM)
&\stackrel{{\rm MF}}{=}&
{1\over{\vol}}\int d\mylolen_{1}^2\,d\mylolen_{2}^2\,
\mydistll(\mylolen_{1}^2)\,
\mydistll(\mylolen_{2}^2)\,
\delta\!\left(\rmsFlDUM_{1}-\mylolen_{1}^2\Scale[1.2]{\mathbbm{1}}\right)\,
\delta\!\left(\rmsFlDUM_{2}-\mylolen_{2}^2\Scale[1.2]{\mathbbm{1}}\right)\,
\delta(\rmsCoDUM),
\\
\distPHY({\rmsFlDUM})
&\stackrel{{\rm MF}}{=}&
\int d\mylolen^2\,\mydistll(\mylolen^2)\,
\delta\!\left(\rmsFlDUM-\mylolen^2\Scale[1.2]{\mathbbm{1}}\right)\,,
\end{eqnarray}
\end{subequations}
where $\Scale[1.2]{\mathbbm{1}}$ is the $\Dim\times\Dim$ identity tensor for the Cartesian space of positions.
Evidently, within this mean-field analysis the thermal fluctuations of the positions of the localized particles are spatially isotropic, and thus fully characterized by the localization-length distribution $\mydistll(\mylolen^2)$.
Moreover, the heterogeneity distribution is insensitive to the separation between the mean positions of the particle pairs.
Furthermore, the localization lengths of the particle pairs are uncorrelated
and the thermal motions of the the particle pairs are uncorrelated, too.
Unsurprisingly, the picture of the amorphous solid state painted by mean field theory is limited. It is necessary to incorporate the effects of order-parameter fluctuations in order to uncover a more refined picture of the statistical mesoscale structure of the amorphous solid state.

In order to go beyond mean-field theory, we now incorporate the effects of $\Omega$-field fluctuations for modes on the gapless branch of excitations (\ie, the Goldstone branch); see  Refs.~\cite{MGXZepl-2007,XMMphd-2008,MGXZpre-2009}.
The structure of these excitations results from the pattern of spontaneous symmetry breaking (\viz, the reduction in symmetry from independent translations of the replicas down to their common translations).
They are readily expressed as deformations of the mean-field expectation value of the replica order parameter that are parametrized by means of a set of $\repsREP$ (not $\repsREP+1$) $\Dim$-dimensional vector fields $\{\udisvec^{\alphaREP}(\zedvec)\}_{\alpha=1}^{\repsREP}$. These fields  depend on the ($\Dim$-dimensional) spatial position $\zedvec$ and are appropriately reminiscent of the displacement fields of a (conventionally replicated) elasticity theory; see Refs.~\cite{MGXZepl-2007,XMMphd-2008,MGXZpre-2009}.
Consistent with the idea that common translations of the replicas remain symmetries of the amorphous solid state, there is one fewer displacement field than there are replicas; this is why $\udisvec^{0}(\zedvec)$ is absent from the theory.
Each of these displacement fields obeys the incompressibility condition
${\rm det}_{D\times D}
\big(\delta_{dd^{\prime}}+
\partial u_{d}^{\alpha}(\zedvec)/
\partial z_{d^{\prime}}^{\phantom\alpha}\big)=1$, 
and thus is associated with a pure shear deformation.
Focusing, just for now, only on the part associated with localized particles, how the displacement fields give rise to a Goldstone-branch deformation of the replica order parameter away from its equilibrium expectation value is specified as follows:
\begin{subequations}
\begin{eqnarray}
\label{eq:unGSdeformed}
&&\replangle\,
\Omega(\khat)\,
\reprangleMF=
\locfrac\,
\mydistlltrans\left(\khat^2\right)
\int{d^{D}z\over{V}}\,
\exp\left({i\sum\nolimits_{\alpha=0}^{\repsREP}\kvec^{\alpha}
\cdot\zedvec}\right),
\\[2pt]
&&\qquad\longrightarrow
\locfrac\,
\mydistlltrans\left(\khat^{2}\right)
\int{d^{D}z\over{V}}\,
\exp\left({i\sum\nolimits_{\alpha=0}^{\repsREP}\kvec^{\alpha}
\cdot\zedvec}\right)\,
\exp\left({i\sum\nolimits_{\alpha=1}^{\repsREP}\kvec^{\alpha}
\cdot\udisvec^{\alpha}(\zedvec)}\right).
\label{eq:GSdeformed}
\end{eqnarray}%
\end{subequations}%

We define the relationship between displacement fields
$\{\udisvec^{\alpha}(\zedvec)\}_{\alpha=1}^{n}$
and their Fourier transforms
$\{\udisvecF^{\alpha}(\qvec)\}_{\alpha=1}^{n}$
\via\ the pair:
\begin{subequations}
\begin{eqnarray}
\udisvecF(\qvec)
&=&
\int{d^{D}z}\,
{\rm e}^{i\qvec\cdot\zedvec}\,
\udisvec(\zedvec),
\\
\udisvec(\zedvec)
&=&
{1\over{\vol}}\sum\nolimits_{\qvec}
{\rm e}^{-i\qvec\cdot\zedvec}\,
\udisvecF(\qvec).
\end{eqnarray}%
\end{subequations}%
Each wave-vector-space displacement field $\udisvecF(\qvec)$ is defined on a (fine) discrete wave-vector lattice with lattice spacing set by the inverse of the (large) real-space system size. 
In view of the existence of gapped excitation branches, the Goldstone-branch description breaks down for Goldstone-branch excitation energies that exceed the lowest-energy gapped excitation energies (\ie, have wave-vectors comparable to or shorter than the inverse of the typical localization length). Therefore, to be internally consistent, we cut off the Goldstone-branch contributions associated with wavelengths shorter than the typical inverse localization length by truncating wave-vector summations at this scale. 

The fluctuations of the displacement fields are governed by an effective elastic Hamiltonian 
$\delta\efe\equiv\delta\efee+\delta\efem$ 
that comprises two contributions. 
The second, $\delta\efem$, which we explain shortly, is induced by the transformation of the functional integration measure from $\Omega$ to $\udisvecF$. 
The first, $\delta\efee\equiv\efee[\,\opfe,\udisvecF\,\big]-\efee[\,\opfe,\zvec\,\big]$, 
is the increase in the replica Hamiltonian~\eqref{eq:Landau-Wilson} when one replaces the mean-field replica order parameter $\opfe$ 
[see Eq.~\eqref{eq:unGSdeformed}] by its Goldstone-distorted counterpart~\eqref{eq:GSdeformed}, as explained, 
\eg, in Ref.~\cite{ref:ZHG-2022arxiv}; to leading order in displacement fields it is given by
\begin{subequations}%
\begin{equation}
\label{eq:elas}
\delta\efee\big[\,\opfe,\udisvecF\,\big]=
-{\pno\over{2\vol}}
\sum_{\underset{(\alpha_{1}\,\ne\,\alpha_{2})}
        {\alpha_{1},\,\alpha_{2}\,=\,1}}^{n}
{1\over{\vol}}\sum_{\qvec}\,
\sdse(\qvec)\,\vert\qvec\vert^{2}\,
\udisvecF^{\alpha_{1}}(\qvec)^{\ast}\cdot
\udisvecF^{\alpha_{2}}(\qvec),
\end{equation}%
where $\sdse(\qvec)$ is a wave-vector-dependent elastic shear modulus. As shown in Ref.~\cite{ref:ZHG-2022arxiv}, $\sdse(\qvec)$ and its long-distance limit $\sdse(\zvec)$ are determined by the dimensionless scaling function $\Sigma(\kappa)$ \via\ 
\begin{eqnarray}%
\sdse(\qvec)
&=&
\sdse(\bm{0})\,
\Sigma(\vert\qvec\vert^{2}/\ctp),
\\
\sdse\big(\bm{0}\big)
&=&
T\,(4\,\ctp^{3}/27),
\end{eqnarray}%
\label{eq:setof}%
\end{subequations}%
where Boltzmann's constant has been set to unity. Note that the distribution of localization lengths $\mydistll$ determines the wave-vector-dependent elastic shear modulus $\sdse(\qvec)$ \via\ $\Sigma(\kappa)$. 

As for the second contribution, $\delta\efem$, this arises from the transformation of the functional integral over $\Omega$ into the functional integral over {$\udisvec$}, the latter integral reaching only the slice (or submanifold) of $\Omega$-space determined by Eq.~\eqref{eq:GSdeformed}. This transformation induces a measure $\sqrt{\det\myIndMet}$ for the $\udisvec$ functional integral. 
Explicitly, the two integration measures are related by 
\begin{equation}
    \int\cdots\mathcal D\Omega=
    \int\cdots\sqrt{\det\myIndMet[\udisvec]}\ \mathcal D\udisvec\,, 
\end{equation} 
where the {\it induced metric\/} $\myIndMet$ is defined \via\  
\begin{equation}
    \myIndMet(\zedvec_2,\alpha_2,d_2;\zedvec_1,\alpha_1,d_1)\equiv
    {\frac{1}{2}}{\sum_{\khat\in\hrs}}\Re
    \fdv{\Omega^*(\khat)}{u^{\alpha_2,d_2}(\zedvec_2)}\fdv{\Omega(\khat)}{u^{\alpha_1,d_1}(\zedvec_1)}\,, 
\end{equation} 
in which each $u^{\alpha,d}(\zedvec)$ labels an individual degree of freedom of $\udisvec$ at 
position $\zedvec$ with replica label $\alpha$ and vector index $d$, and $\det$ indicates a Fredholm determinant; 
for a simple illustrative example, see Ref.~\cite{Toymetric}. 
It is straightforward to compute $\myIndMet$ starting from Eq.~\eqref{eq:GSdeformed}; however, computing the Fredholm determinant is more subtle, 
because of the need for the displacement field to obey the incompressibility constraint. 
We shall publish details of these computations elsewhere~\cite{ref:ZZG-IM}. 
To understand the effect of the induced measure $\sqrt{\det\myIndMet}$, we estimate its contribution to $\delta\efem\,$. 
For small spatial gradients of $\udisvec$, its leading behavior can be shown to have the form 
\begin{equation}
\delta\efem
\approx
\frac{\pno}{2\vol}
\sum_{\alpha_1,\alpha_2=1}^n
\frac{1}{\vol}\sum_\qvec
\sdsm^{\alpha_1,\alpha_2}(\qvec)\,\vert\qvec\vert^{2}\,
\udisvecF^{\alpha_{1}}(\qvec)^{\ast}\cdot
\udisvecF^{\alpha_{2}}(\qvec)\,, 
\end{equation}%
in terms of a particular function $\sdsm$, which is an 
induced-measure--generated contribution to the scale-dependent elastic shear modulus. 
We focus on the long-wavelength behavior of  
$\sdsm^{\alpha_1,\alpha_2}(\qvec)$ 
so that we may contrast it with the long-wavelength behavior of 
$\sdse(\qvec)$ as, together, they characterize the macroscopic linear elasticity. 
Thus, we consider 
\begin{equation}
    \sdsm^{\alpha_1,\alpha_2}(\qvec)\big\vert_{\qvec\to\zvec}=
    \frac{T}{N}
    \sum_{\kvec}^{\xi_{\rm typ}\abs{\kvec}\leq 1}
    \myScaleFn^{\alpha_1,\alpha_2}(\opfe,\kvec)\,, 
\end{equation} 
where $\xi_{\rm typ}$ is the typical localization length which scales as $\xi_{\rm typ}\sim\xi_0/\sqrt{\tau}$ at mean-field level, and $\myScaleFn(\opfe,\kvec)$ 
is a certain wave-vector-dependent function that also depends on information contained in 
the equilibrium value of the order-parameter field $\opfe$. 
A detailed quantitative analysis~\cite{ref:ZZG-IM} ensures that $\myScaleFn$ is at most of order unity.
As a result, in the small wave-vector limit 
the induced-measure--generated shear modulus scales as  
\begin{equation}\label{eq:elas-induced-measure-estimate}
    \sdsm(\zvec)\sim
    \frac{TV}{N\xi_0^\Dim}\,\tau^{\Dim/2}\,. 
\end{equation} 
In Eq.~\eqref{eq:elas-induced-measure-estimate}, the factor $V/N\xi_0^\Dim$ determines the extent to which the polymer coils typically overlap one another. 
A comparison of the factor $\tau^{3}$ in $\sdse(\zvec)$ and the factor $\tau^{\Dim/2}$ in $\sdsm(\zvec)$ shows that for $\Dim>6$ the free-energy contribution to the shear modulus dominates the induced-measure contribution, at least asymptotically close to the amorphous solidification critical point. At six dimensions, the induced-measure contribution to the shear modulus is no larger than the free-energy contribution, although it could potentially modify both the macroscopic and the (dimensionless) scale-dependent shear moduli.
We cannot be certain that the dominance of the free-energy contribution continues below six dimensions because the scaling properties of the amorphous solid state below six dimensions (which play a role in determining both contributions) are unknown at this time. 
However, we shall limit our scope to the regime in which the dominance by the free-energy contribution holds (\ie, $\Dim>6$), and assume that this dominance holds not only in the long-wavelength limit, but also for nonzero wave-vectors. 

To average over the displacement field fluctuations and thus incorporate the Goldstone branch of excitations, we begin with the weight $\exp\left({-\delta\efee}\right)$ 
(having omitted the contribution $\delta\efem$, in line with the discussion given above), denoting the resulting averages by $\GBlangle\cdots\GBrangle$. 
At this stage, we are already retaining terms only to quadratic order in the displacement fields in $\delta\efee\,$. Correspondingly, we also approximate the incompressibility constraint 
${\rm det}_{D\times D}\big(\delta_{d\overbar{d}}+
\partial u_{d}^{\alpha}(\zedvec)/
\partial z_{\overbar{d}}^{\phantom\alpha}\big)=1$, 
which reflects the strong inter-particle repulsion, 
replacing it by the linearized form 
$\partial u_{d}^{\alpha}(\zedvec)/\partial z_{d}^{\phantom\alpha}=0$, 
valid for small displacement-gradients. 
Equipped with these simplifications, it is straightforward to evaluate the average of $\udisvec^{\alpha}$-dependent functions. 
In particular, displacement-field correlator is given by 
\begin{subequations}
\begin{equation}
\GBlangle
u_{\cartD}^{\alpha}(\zedvec_{1})\,
u_{\cartDbar}^{\beta}(\zedvec_{2})\GBrangle=
\big(\delta^{\alpha\beta}+\dfrac{1}{1-\repsREP}\big)
\dfrac{\temp\vol}{\pno}\dfrac{1}{\sdse\big(\bm{0}\big)}
\Ucorten_{\cartD\cartDbar}(\zedvec_{1}-\zedvec_{2})\,,
\label{eq:Ucorpartone}
\end{equation}
where the incompressible elastic Green tensor $\Ucorten$ is given by
\begin{eqnarray}
\Ucorten_{\cartD\cartDbar}
(\bm{\zedvec})=
{1\over{\vol}}
    \sum_{\qvec}^{\xi_{\rm typ}\abs{\qvec}\leq 1}
\dfrac{\rm {e}^{-i\qvec\cdot\zedvec}}{\vert\qvec\vert^{2}}\,
\dfrac{1}{\Sigma(\vert\qvec\vert^{2}/\ctp)}\,
\Big(
\delta_{\cartD\cartDbar}-
\dfrac{
(\qvec)_{\cartD}^{\myalign}\,
(\qvec)_{\overbar{{\cartD}^{\myalign}}}}{\vert\qvec\vert^{2}}
\Big)\,.
\label{eq:Ucorparttwo}%
\end{eqnarray}%
\label{eq:Ucor}%
\end{subequations}%
$\Ucorten$, which features the (dimensionless) scale-dependent shear modulus $\Sigma$, characterizes the displacement at location $\bm{\zedvec}$ in response to a displacement at the origin. It plays a key role in determining the heterogeneity distribution. Its anisotropy results from incompressibility. 
It is long ranged, in the sense that it decays at large distances algebraically for $\Dim>2$; for $\Dim=2$ it diverges logarithmically at distances much larger than the typical localization length; see, Ref.~\cite{ref:GMZcalc}.
We note that $\Ucorten(\bm{0})$ is proportional to the $D\times D$ identity matrix $\Scale[1.2]{\mathbbm{1}}$, from which it is straightforward to show that 
\begin{subequations}\label{eq:Ucorzero}
\begin{eqnarray}
\Ucorten(\bm{0})
&=&
\Ucorsca\,\Scale[1.2]{\mathbbm{1}}\,,
\\
\Ucorsca
&\equiv&
\dfrac{\Dim-1}{\Dim}\,
{1\over{\vol}}\sum_{\qvec}^{\xi_{\rm typ}\abs{\qvec}\leq 1}
\dfrac{1}{\vert\qvec\vert^{2}}\,
\dfrac{1}{\Sigma(\vert\qvec\vert^{2}/\ctp)}\,.
\end{eqnarray}%
\end{subequations}
Importantly, for $\Dim=2$, $\Ucorsca$ diverges logarithmically in the long-distance cutoff (\ie, the system size); see, Ref.~\cite{ref:GMZcalc}.
The displacement-field correlator~\eqref{eq:Ucorpartone} enables us to determine the replica order parameter and replica correlator, averaged over the Goldstone sector. 
For example, the part of the replica correlator that corresponds to the case where both particles are localized reads  
\begin{subequations}
\begin{eqnarray}
&&
\replangle\,
\Omega(\khat_{1})^{\ast}\,
\Omega(\khat_{2})
\reprangleGB\Big|_\text{loc-loc}
\nonumber\\[4pt]
&&\,=\locfrac\,
\mydistlltrans(\khat_{1}^{2})
\int{d^{D}z_{1}\over{V}}\,
\exp\big({-i\sum\nolimits_{\alpha=0}^{\repsREP}\kvec_{1}^{\alpha}
\cdot\zedvec_{1}}\big)\,
\locfrac\,
\mydistlltrans(\khat_{2}^{2})
\int{d^{D}z_{2}\over{V}}\,
\exp\big({i\sum\nolimits_{\alpha=0}^{\repsREP}\kvec_{2}^{\alpha}
\cdot\zedvec_{2}}\big)
\nonumber\\[4pt]
&&\,\qquad\qquad\times
{\GBlangle}
\exp\big({-i\sum\nolimits_{\alpha=1}^{\repsREP}\kvec_{1}^{\alpha}
\cdot\udisvec^{\alpha}(\zedvec_{1})}\big)
\exp\big({i\sum\nolimits_{\alpha=1}^{\repsREP}\kvec_{2}^{\alpha}
\cdot\udisvec^{\alpha}(\zedvec_{2})}\big)
\GBrangle
\nonumber\\[4pt]
&&\,
=\locfrac^2\,
\delta_{\sum\nolimits_{\alphaREP=0}^{\repsREP}{\kvec_{1}^{\alphaREP}},\;
\sum\nolimits_{\alphaREP=0}^{\repsREP}{\kvec_{2}^{\alphaREP}}}\,
\mydistlltrans(\khat_{1}^{2})\mydistlltrans(\khat_{2}^2)
\int{d^{\Dim}z\over{V}}\,
\exp\left({i\sum\nolimits_{\alpha=0}^{\repsREP}\kvec_{1}^{\alpha}
\cdot\zedvec}\right)\corrker(\khat_1,\khat_2;\zedvec)\,, 
\label{eq:repcorlocloc}
\end{eqnarray}%
in which for convenience we have introduced the replica correlation kernel $\corrker(\khat_1,\khat_2;\zedvec)$, defined \via\ 
\begin{eqnarray}
&&
\corrker(\khat_1,\khat_2;\zedvec_{1}-\zedvec_{2})
\equiv
\GBlangle
\exp\big({-i\sum\nolimits_{\alpha=1}^{\repsREP}\kvec_{1}^{\alpha}
\cdot\udisvec^{\alpha}(\zedvec_{1})}\big)
\exp\big({i\sum\nolimits_{\alpha=1}^{\repsREP}\kvec_{2}^{\alpha}
\cdot\udisvec^{\alpha}(\zedvec_{2})}\big)
\GBrangle
\nonumber
\\[4pt]
&&
\qquad\quad=
\exp\Bigg(-\dfrac{1}{2}\dfrac{\vol}{\pno}\dfrac{\temp}{\sdse\big(\bm{0}\big)}
\sum_{\alpha,\beta=1}^{\repsREP}\Big(\delta_{\alpha\beta}+\dfrac{1}{1-\repsREP}\Big)
\nonumber\\[4pt]
&&\,\quad\times
\Big[\Ucorten_{\cartD\cartDbar}(\bm{0})\,
(\kvec_{1})_{\cartD}^{\alpha}\,
(\kvec_{1})_{\cartDbar}^{\beta}+
\Ucorten_{\cartD\cartDbar}(\bm{0})\,
(\kvec_{2})_{\cartD}^{\alpha}\,
(\kvec_{2})_{\cartDbar}^{\beta}-
2\Ucorten_{\cartD\cartDbar}(\zedvec_{1}-\zedvec_{2})\,
(\kvec_{1})_{\cartD}^{\alpha}\,
(\kvec_{2})_{\cartDbar}^{\beta}\Big]\Bigg)\,. 
\label{eq:repcorrkerdef}
\end{eqnarray}%
\end{subequations}%
We specialize to the notation $\corrker(\khat)$ when one or other of the wave-vector arguments vanishes, in which case $\corrker(\khat)$ loses $\zedvec$-dependence. 
Thus, the full replica order parameter and replica correlator take the respective forms 
\begin{subequations}
\begin{eqnarray}
\replangle\,
\Omega(\khat)\,
\reprangleGB
&&=(1-\locfrac)\,\delta_{\zhat,\,\khat}\,+
\locfrac\,
\delta_{\sum\nolimits_{\alphaREP=0}^{\repsREP}{\kvec^{\alphaREP}},
\,\zvec}\,
\mydistlltrans(\khat^2)\,\corrker(\khat)\,,
\label{eq:repop}
\\[4pt]
\replangle\,
\Omega(\khat_{1})^{\ast}\,
\Omega(\khat_{2})
\reprangleGB
&&\,=
\locfrac^2\,
\delta_{\sum\nolimits_{\alphaREP=0}^{\repsREP}{\kvec_{1}^{\alphaREP}},\;
\sum\nolimits_{\alphaREP=0}^{\repsREP}{\kvec_{2}^{\alphaREP}}}\,
\mydistlltrans(\khat_{1}^{2})\mydistlltrans(\khat_{2}^2)
\int{d^{\Dim}z\over{V}}\,
\exp\left({i\sum\nolimits_{\alpha=0}^{\repsREP}\kvec_{1}^{\alpha}
\cdot\zedvec}\right)\,\corrker(\khat_1,\khat_2;\zedvec) 
\nonumber\\[4pt]
&&\,\quad+(1-\locfrac)\,\locfrac\,\delta_{\zhat,\,\khat_{1}}\,
\delta_{\sum\nolimits_{\alphaREP=0}^{\repsREP}{\kvec_{2}^{\alphaREP}},\,\zvec}\,
\mydistlltrans(\khat_{2}^2)\,\corrker(\khat_2)
\nonumber\\[4pt]
&&\,\quad+(1-\locfrac)\,\locfrac\,\delta_{\zhat,\,\khat_{2}}\,
\delta_{\sum\nolimits_{\alphaREP=0}^{\repsREP}{\kvec_{1}^{\alphaREP}},\,
\zvec}\,\mydistlltrans(\khat_{1}^2)\,\corrker(\khat_1)
\nonumber\\[4pt]
&&\,\quad+(1-\locfrac)^2\,\delta_{\zhat,\,\khat_{1}}\,\delta_{\zhat,\,\khat_{2}}\,.
\label{eq:repcor}
\end{eqnarray}%
\end{subequations}%
Strictly speaking, 
$\replangle\,\Omega(\khat)\reprangleGB$ and $\replangle\,\Omega(\khat_{1})^{\ast}\,\Omega(\khat_{2})\reprangleGB$ 
vanish for replicated wave-vectors lying in the one-replica sector, owing to the vanishing of the Goldstone distortions of the order-parameter field in this sector together with the incompressibility of the displacement fields. However, as mentioned above, we have not handled this incompressibility exactly in our averaging over the Goldstone fields, but only in the linearized form. As a result, $\replangle\,\Omega(\khat_{1})^{\ast}\,\Omega(\khat_{2})\reprangleGB$ does not exactly vanish in the one-replica sector (although $\replangle\,\Omega(\khat)\reprangleGB$ does).

Having determined the replica order parameter and replica correlator, the methodology given in Eqs.~\eqref{eq:REPcorform} and \eqref{eq:distmatching} brings us to the distribution of single-particle position-fluctuation tensors $\distPHY$ and the heterogeneity distribution 
$\ProbCharsLarge\,$:
\begin{subequations}
\label{eq:distres}
\begin{eqnarray}
&&
\ProbCharsLarge(\MUvecSep,
\rmsFlDUM_{1},\rmsFlDUM_{2},\rmsCoDUM)
=
\int d\mylolen_{1}^2\,d\mylolen_{2}^2\,
\mydistll(\mylolen_{1}^2)\,
\mydistll(\mylolen_{2}^2)\,
\delta\!\left(
\rmsFlDUM_{1}-
\mylolen_{1}^2\Scale[1.2]{\mathbbm{1}}-
\shiftll^2\Scale[1.2]{\mathbbm{1}}\right)\,
\delta\!\left(
\rmsFlDUM_{2}-
\mylolen_{2}^2\Scale[1.2]{\mathbbm{1}}-
\shiftll^2\Scale[1.2]{\mathbbm{1}}\right)
\nonumber
\label{eq:distresA}
\\
&&\qquad\quad\times
\int{d^{\Dim}z\over{\vol}}\;
\delta\!
\left(
\rmsCoDUM-
\shiftll^2\,
\Ucorsca^{-1}\,\Ucorten(\zedvec)
\right)\,
\nonumber\\
&&\qquad\quad\times
\big(4\pi\shiftll^2\big)^{-\frac{\Dim}{2}}\,
\big[\det\swten(\zedvec)\big]^{-\frac{1}{2}}
\exp\left(
{-\dfrac{1}{4\shiftll^2}}\,
\big(\swten(\zedvec)^{-1}\big)_{\cartD\cartDbar}\,
(\zedvec-\MUvecSep)_{\cartD}\,(\zedvec-\MUvecSep)_{\cartDbar}\right),
\\
\noalign{\medskip}
&&
\distPHY({\rmsFlDUM})=
\int d\mylolen^2\,\mydistll(\mylolen^2)\,
\delta\!\left(
\rmsFlDUM-
\mylolen^2\Scale[1.2]{\mathbbm{1}}-
\shiftll^2\Scale[1.2]{\mathbbm{1}}\right)
=\int d\mylolen^2\,\myshiftdistll(\mylolen^2)\,
\delta\!\left(\rmsFlDUM-\mylolen^2\Scale[1.2]{\mathbbm{1}}\right)\,,
\label{eq:distresB}
\end{eqnarray}
\end{subequations}
where the shift length $\shiftll$,
covariance matrix $\swten$
and shifted distribution $\myshiftdistll$
are defined \via\
\begin{subequations}
\begin{eqnarray}
\shiftll^2 \,&\equiv&\,\dfrac{\temp\vol}{\pno}
\dfrac{\Ucorsca}{\sdse(\bm{0})},
\label{eq:distresTermA}
\\
\swten(\zedvec)\,&\equiv&\,\Scale[1.2]{\mathbbm{1}}-
\Ucorsca^{-1}\,\Ucorten(\zedvec)\,,
\label{eq:distresTermB}
\\
\myshiftdistll(\mylolen^2)\,
&\equiv&\,
\mydistll(\mylolen^2-\shiftll^2)\,.
\label{eq:distresTermC}
\end{eqnarray}
\end{subequations}
The heterogeneity distribution $\ProbCharsLarge$ -- which extends the position-fluctuation distribution $\distPHY$ from one-particle to two-particle localization characteristics -- is the core result of this Paper. 
Below, we unpack $\ProbCharsLarge$ to see what it says about the statistical mesoscale structure of the amorphous solid state, but first we address what is learned from $\distPHY$.
\hfil\break\noindent
(i)~Equation~\eqref{eq:distresB} for $\distPHY$ shows that the single-particle thermal position-fluctuation tensors $\rmsFlDUM$, which at the mean-field level were isotropic and characterized by the distribution $\mydistll$ of squared localization lengths, remain isotropic upon inclusion of Goldstone excitations; however, the weight of the squared localization-length distribution is uniformly shifted to longer lengths, by the square of the length $\shiftll$, to become the distribution $\myshiftdistll$. 
As the discussion below Eqs.~\eqref{eq:Ucorzero} indicates, for $\Dim>2$ this shift is finite, whereas at $\Dim=2$ it diverges with the system size. Dimensional analysis shows that for $\Dim>6$ (which is within the scope of the present Paper), the shift length $\shiftll$ is much smaller the bare (\ie, mean-field) typical localization length $\xi_{\rm typ}$. 
Equation~\eqref{eq:distresTermA} shows that the shift length is determined \via\ the scale-dependent shear modulus, which in turn is dictated by the (scaled, mean-field) distribution of squared localization lengths. (Thus, the mean-field distribution encodes instructions for how it is to be renormalized by Goldstone excitations.)\thinspace\
\hfil\break\noindent 
\def\viz{{\it viz\/}.}
(ii)~Equation~\eqref{eq:distresA} gives -- within the Goldstone-excitation framework -- the heterogeneity distribution $\ProbCharsLarge$. It shows that within this framework the localization lengths of particle pairs are uncorrelated with one another, regardless of the separation of their mean positions $\MUvecSep$.
What {\it does\/}, however, depend on the separation -- within the Goldstone framework -- is the distribution of the tensor $\rmsCoDUM$ (\ie, the correlator between the thermal position fluctuations of a pair of particles away from their mean positions). The recipe for this distribution is as follows. 
Start with the given separation $\MUvecSep$ of the mean particle-positions. Associate with $\MUvecSep$ a sharp value of the correlator $\rmsCoDUM$, \viz,
$\shiftll^2\,\Ucorsca^{-1}\,\Ucorten(\MUvecSep)$. 
Now smear $\MUvecSep$ (and hence $\rmsCoDUM$) by making random departures from $\MUvecSep$ that are managed by the quasi-Gaussian random variable $\zedvec$. These departures have their anisotropy governed by the covariance tensor $\swten(\zedvec)$, and their scale set by $\swten(\zedvec)$ and the shift length $\shiftll$, both of which are indirectly controlled by the scale-dependent shear modulus. (We use the term {\it quasi-Gaussian\/} to indicate the dependence of the covariance on $\zedvec$.)\thinspace\ The anisotropy of the position-fluctuation correlations $\rmsCoDUM$ is inherited from the anisotropy of the elastic Green function $\Ucorten$, which itself is a consequence of the incompressibility of the displacement fluctuations $\{\udisvec^{\alpha}(\zedvec)\}_{\alpha=1}^{n}$ that results from inter-particle repulsion. 
As for the impact of the scale of the particle separation $\MUvecSep$ on how $\rmsCoDUM$ is distributed, this is in accordance with physical expectations. When the separation is asymptotically large 
(compared with $\xi_{\rm typ}$), 
the distribution is dominated by correspondingly large values of $\zedvec$ and, therefore, by vanishingly small values of $\rmsCoDUM$, owing to the vanishing of $\Ucorten$ at large distances. Thus, particle pairs whose mean positions are widely separated have essentially uncorrelated position-fluctuation correlations. 
On the other hand, as the separation is reduced, the distribution of position-fluctuation correlations sharpens towards a deterministic value of $\shiftll^2\,
\Ucorsca^{-1}\,\Ucorten(\MUvecSep)$, owing to the reduction in smearing. Accompanying this sharpening is an increase in the typical strength of $\rmsCoDUM$, with the scale of $\rmsCoDUM$ reaching $\shiftll^2$ as the separation of the mean positions is reduced to much smaller than $\xi_{\rm typ}$. As mentioned above, within the scope of the present Paper (\ie, $\Dim>6$), $\shiftll$ is much smaller than the typical localization length $\xi_{\rm typ}\,$, and thus the particle-pair correlations are weak compared with the position fluctuations of the individual particles. 

What should one anticipate as the consequences of going beyond the Goldstone-excitation framework by also incorporating gapped sectors of excitations? 
The most prominent consequence would be that the heterogeneity distribution $\ProbCharsLarge$ acquires correlations between the single-particle position-fluctuation tensors $\rmsFlDUM_{1}$ and $\rmsFlDUM_{2}$. 
For mean-position separations $\MUvecSep$ much smaller than $\xi_{\rm typ}$, there should be essentially perfect correlation between $\rmsFlDUM_{1}$ and $\rmsFlDUM_{2}$; on the other hand, for much larger separations $\rmsFlDUM_{1}$ and $\rmsFlDUM_{2}$ should become essentially uncorrelated. 
On top of this, one should anticipate a more refined result for the fluctuation-corrected distribution of (squared) localization lengths. It will not merely undergo the rigid shift necessarily dictated by the structure of Goldstone excitations, as given in Eq.~\eqref{eq:distresTermC}. Instead, the change in the dependence on $\mylolen^2$ will be more generic, reflecting the detailed structure of the gapped excitations. 
\section{Concluding remarks
\label{sec:conclude}}
The replica-statistical field theory governing the phase transition to and equilibrium properties of the emergent amorphous solid state is universal, in the sense that its structure is determined solely \via\ symmetry and length-scale arguments. 
We have examined this field theory with a view to extracting the (correspondingly universal) statistical information it encodes about the amorphous solid state -- information, which we refer to as the mesoscale heterogeneity of the state, that arises as a result of the random architecture of the underlying medium and the consequent random point-to-point variation of its emergent properties. 
This information takes the form of a joint probability distribution for the spatial heterogeneity of the thermal motion of the constituents of the amorphous-solid-forming system. Covered by the distribution are the following characteristics of the thermal motion: the R.M.S.~values of the fluctuations of the positions of the constituents about their (random) mean values, and the correlations between these position fluctuations. The distribution quantifies how these random characteristics are jointly distributed, and how their distribution varies with the separation of the mean positions of pairs of constituents. This information is encoded in and extracted from the correlators of the amorphous solidification field theory. 

We have analyzed increasingly accurate approximations to the order parameter and correlators, ranging from mean-field theory to the incorporation of gapless elastic-displacement (\ie, Goldstone) fluctuations. Hence, we have obtained increasingly accurate descriptions of the mesoscale heterogeneity. We have paid attention to the role of the induced measure that arises when the (complete) functional integral over the fluctuating order-parameter field is approximated by means of a (partial) functional integral over elastic displacement fields. 

Several key themes remain for future examination. These include the quantitative analysis of the impact of the gapped (\ie, non-Goldstone) sector of field fluctuations on the heterogeneity distribution and the development of a model-independent scaling theory for the heterogeneity distribution that would allow for the consequences of critical fluctuations. 
\begin{acknowledgments}
This work was performed in part at the Aspen Center for Physics, which is supported by National Science Foundation grant PHY-2210452.
\end{acknowledgments}

{\raggedright 

} 


\begin{thebibliography}{99}
\bibitem{note:cor2op}
The reduction from the physical correlator to the physical order parameter is effected 
in Eq.~\eqref{eq:twoPHYcorsmom} by setting to zero one or other of the the wave-vector sets
$\{\qvec_{1}^{\alphaPHY}\}_{\alphaPHY=1}^{\repsPHY}$ or 
$\{\qvec_{2}^{\alphaPHY}\}_{\alphaPHY=1}^{\repsPHY}$, 
and in Eq.~\eqref{eq:twoPHYcorsRS} integrating over one or other of the position-vector sets 
$\{\rvec_{1}^{\alphaPHY}\}_{\alphaPHY=1}^{\repsPHY}$ or 
$\{\rvec_{2}^{\alphaPHY}\}_{\alphaPHY=1}^{\repsPHY}\,$.
\bibitem{ref:cumulant}
This assumption is justified \via\ a cumulant expansion, in which corrections to the Gaussian form are negligibly small, at least near to the solidification transition (where localization is weak). Short-distance correlations resulting, say, from inter-particle repulsion are not apparent in the mesoscopic-to-macroscopic regime of length-scales under consideration here.
\bibitem{Toymetric}
As an elementary example of an induced measure, consider a unit sphere $(\theta,\varphi)$ embedded in a Euclidean space $(x,y,z)$. The Jacobian matrix $J$ for the transformation from $(x,y,z)$ to $(\theta,\varphi)$ is given by 
$$J=\mqty(\pdv*{x}{\theta}&\pdv*{x}{\varphi}\\\pdv*{y}{\theta}&\pdv*{y}{\varphi}\\\pdv*{z}{\theta}&\pdv*{z}{\varphi})=\mqty(\cos\theta\cos\varphi&-\sin\theta\sin\varphi\\\cos\theta\sin\varphi&\sin\theta\cos\varphi\\-\sin\theta&0)\,.$$
The sphere inherits an induced metric $g$ from the ambient Euclidean space, which can be computed using the Jacobian as 
$$g=J^\dagger J=\mqty(1&0 \\
         0&\sin^{2}\theta)\,.$$
The induced integration measure can then be expressed in terms of the induced metric $g$ \via 
$$\delta(1-x^2-y^2-z^2)\,\dd x\,\dd y\,\dd z\propto\sqrt{\det g}\,\dd\theta\,\dd\varphi=\sin\theta\,\dd\theta\,\dd\varphi\,,$$
where the $\delta$ function on the left-hand side serves to constrain the integration to the appropriate subspace, \ie, the surface of the unit sphere.
\end{thebibliography}
\end{document}